\documentclass[twocolumn]{aastex63}

\graphicspath{{./}{figures/}}

\received{}
\revised{}
\accepted{}
\submitjournal{ApJ}

\shorttitle{Chemistry of the apolar ice layer}
\shortauthors{R. Mart\'in-Dom\'enech et al.}
\begin{document}

\title{Exploring the chemistry induced by energetic processing of the H$_2$-bearing, CO-rich apolar ice layer}

\correspondingauthor{Rafael Mart\'in-Dom\'enech}
\email{rafael.martin\_domenech@cfa.harvard.edu}

\author{Rafael Mart\'in-Dom\'enech}
\affil{Center for Astrophysics $|$ Harvard \& Smithsonian \\
60 Garden St., Cambridge, MA 02138, USA}

\author{Pavlo Maksiutenko}
\affil{Center for Astrophysics $|$ Harvard \& Smithsonian \\
60 Garden St., Cambridge, MA 02138, USA}

\author{Karin I. \"Oberg}
\affil{Center for Astrophysics $|$ Harvard \& Smithsonian \\
60 Garden St., Cambridge, MA 02138, USA}

\author{Mahesh Rajappan}
\affil{Center for Astrophysics $|$ Harvard \& Smithsonian \\
60 Garden St., Cambridge, MA 02138, USA}

\begin{abstract}

Interstellar ice mantles on the surfaces of dust grains 
are thought to have a bi-layered structure, with a H$_2$O-rich polar layer, covered by a CO-rich apolar layer that probably harbors H$_2$ and other volatiles such as N$_2$. 
In this work, we explore the chemistry induced by 2 keV electrons and Ly-$\alpha$ photons in H$_2$:CO:$^{15}$N$_2$ ice analogs of the CO-rich layer when exposed to similar fluences to those expected from the cosmic-ray-induced secondary electrons and UV photons during the typical lifetime of dense clouds.
Six products were identified upon 2 keV electron irradiation:  
CO$_2$, 
C$_2$O (and other carbon chain oxides), 
CH$_4$, 
H$_2$CO, 
H$_2$C$_2$O, 
and H$^{15}$NCO. 
The total product abundances  
corresponded to 5$-$10\% of the initial CO molecules exposed to the electron irradiation. 
Ly-$\alpha$ photon irradiation delivered 1$-$2 orders of magnitude lower yields with  a similar product branching ratio, which may be due to the low UV-photon absorption cross-section 
of the ice sample at this wavelength. 
%
%
%
%
Formation of additional N-bearing species, namely C$_2$$^{15}$N$_2$ and $^{15}$NH$_3$, was only observed in the absence of H$_2$ and CO molecules, respectively, suggesting that reactants derived from H$_2$ and CO molecules preferentially react with each other instead of with $^{15}$N$_2$ and its  dissociation products. 
%
In summary, ice chemistry induced by energetic processing of the CO-rich apolar ice layer provides alternative formation pathways for several species detected in the interstellar medium, 
including some related to the complex organic molecule chemistry. 
Further quantification of these pathways will help astrochemical models to
constrain   
their relative contribution to the interstellar budget of, especially, the organic species H$_2$CO 
and HNCO.

\end{abstract}


\section{Introduction} \label{sec:intro}

In the interior of dense molecular clouds, interstellar ice mantles are formed on the surfaces of dust grains in two phases, leading to a bi-layered structure. 
A H$_2$O-rich polar ice layer that includes other molecules formed through grain surface chemistry 
such as CO$_2$, CH$_4$, or NH$_3$,    
is covered by a CO-rich layer that probably contains other volatile species accreted from the gas-phase such as N$_2$ and O$_2$
\citep{boogert15}.  
Molecular hydrogen (H$_2$) is the most abundant molecule in the universe \citep{wooden04}. 
Even though it is a very volatile species, the presence of H$_2$ in the interstellar ice mantles have been considered due to its large abundance in the interstellar medium \citep[ISM, see, e.g.,][]{sandford93,kristensen11}. %
%
\citet{chuang18} recently suggested that under the conditions found in the innermost regions of the dense clouds  
(n $\ge$ 10$^4$ cm$^{-3}$, T $<$ 20 K),  
H$_2$ could co-deposit along with CO molecules during the CO accretion phase of ice mantle formation.

Interstellar ice mantles are characterized by a rich and diverse chemistry that can result in a wealth of complex organic molecules \citep[COMs, see, e.g.,][]{garrod13}, the proposed organic precursors of more complex prebiotic species. 
These COMs are incorporated into protostellar and protoplanetary disks during star formation, and detected in Solar System comets \citep[see, e.g.,][]{goesmann15,altwegg15}. 
%
%
Interstellar ice chemistry has been experimentally studied in the laboratory under astrophysically relevant conditions for a few decades \citep[see, e.g.,][]{gerakines96}. 
Most studies focus either 
on the H-atom addition reactions to molecules in the 
H$_2$O-rich \citep{qasim18,qasim19b,chuang20,qasim20}
or the CO-rich  
\citep{watanabe02,watanabe08,fuchs09,hama13,linnartz15,fedoseev15a,fedoseev17,chuang16,chuang17,qasim19a,qasim19c} ice layers; 
or on the chemistry induced by energetic processing of the H$_2$O-rich layer \citep[see the review in][]{obergrev}. 
The energetic processing of ice mantles in the interior of dense clouds is driven by the secondary keV electrons and the secondary UV field produced by the interaction of the cosmic rays with the ice molecules \citep{bennett05}, and the gas-phase H$_2$ molecules \citep{cecchi92,shen04}, respectively. 
%

Only two works have recently addressed the photon-induced chemistry of binary ice mixtures analog to the CO-rich ice layer.  
Vacuum-ultraviolet (VUV) photon irradiation 
of $^{13}$CO:N$_2$ ices only led to the formation of CO$_2$ molecules \citep{hector19},  
while carbon chain oxides and nitrogen oxides were also formed when the CO:N$_2$ ice was irradiated with protons    
\citep{sicilia12}. 
Photon irradiation 
of H$_2$:CO ice samples 
led to the formation of CO$_2$ and H$_2$CO molecules (a COM precursor), along with unreacted HCO$^.$ radicals \citep{chuang18}.  
This suggests that energetic processing of the CO-rich layer may contribute to the formation of COMs if H$_2$ molecules are also present.
%
In this work, we explore the chemistry induced by 
both 2 keV electron and Ly-$\alpha$ photon irradiation of H$_2$:CO:$^{15}$N$_2$ ice analogs of the CO-rich apolar ice layer in the cold, innermost regions of dense clouds. 
We use the experimental results to elucidate the contribution of this layer to the chemistry of the ISM, and the interstellar budget of organic species in particular.   
The experimental setup is described in Sect. \ref{sec:exp}, and the resulting chemistry is presented in Sect. \ref{sec:results}. The implications of the most important results are discussed in Sect. \ref{sec:disc}. Finally, the conclusions are summarized in Sect. \ref{sec:conc}.

\section{Methods}\label{sec:exp}

\begin{deluxetable*}{cccccccc}
\tablecaption{Summary of the experiments simulating the energetic processing of the apolar ice layer at 4 K.\label{table_exp}}
\tablehead{
\colhead{Exp.} & \colhead{Ice comp.} & \colhead{Composition ratio} & \multicolumn{2}{c}{Ice thickness (ML)$^{a}$}  &  \colhead{Irradiation} & \multicolumn{2}{c}{Irrad. energy ($\times$ 10$^{18}$ eV)} \\
& & & \colhead{Total} & \colhead{Processed$^{b}$} &  & \colhead{Incident} & \colhead{Absorbed}
}
\startdata
1 & H$_2$:CO:$^{15}$N$_2$ & 1.8:1.0:1.0 & 1500 & 380 & 2 keV e$^-$ & 1.3 & 1.3 \\
2 & H$_2$:CO:$^{15}$N$_2$ & 1.6:1.0:1.0 & 1600 & 380 & 2 keV e$^-$ & 1.3 & 1.3 \\
3 & H$_2$:CO:$^{15}$N$_2$ & 1.6:1.0:1.1 & 2400 & 380 & 2 keV e$^-$ & 1.3 & 1.3\\
4 & H$_2$:CO:N$_2$ & 1.5:1.1:1.0 & 1400 & 380 & 2 keV e$^-$ & 1.5 & 1.5\\ 
5 & H$_2$:$^{13}$CO:$^{15}$N$_2$ & 1.1:1.0:1.0 & 2700 & 380 & 2 keV e$^-$ & 1.5 & 1.5\\
6 & H$_2$:C$^{18}$O:$^{15}$N$_2$ & 1.4:1.0:1.2 & 2800 & 380 &2 keV e$^-$ & 1.3 & 1.3\\
7 & D$_2$:$^{13}$CO:$^{15}$N$_2$ & 1.1:1.1:1.0 & 2400 & 380 & 2 keV e$^-$ & 1.3 & 1.3\\
\hline
8 & H$_2$:CO:$^{15}$N$_2$ & 1.7:1.0:1.0 & 1900 & 1900 & Ly-$\alpha$ & 1.3 & 0.065\\ 
\hline
\hline
9 & H$_2$:CO & 1.5:1.0 & 2100 & 420 & 2 keV e$^-$ & 1.3 & 1.3\\
\hline
10 & H$_2$:CO & 1.7:1.0 & 1700 &1700 & Ly-$\alpha$ & 1.4 & 0.078\\ 
\hline
\hline
11 & CO:$^{15}$N$_2$ & 1.0:1.0 & 2300 & 220 & 2 keV e$^-$ & 1.3 & 1.3 \\
\hline
12 & CO:$^{15}$N$_2$ & 1.0:1.0 & 1400 & 1400 & Ly-$\alpha$ & 1.4 & 0.098\\ 
\hline
\hline
13 & H$_2$:$^{15}$N$_2$ & 2.1:1.0 & 900 &460 & 2 keV e$^-$ & 1.3 & 1.3 \\
\hline
14 & H$_2$:$^{15}$N$_2$ & 2.3:1.0 & 1900 & 1900 & Ly-$\alpha$ & 1.3 & 5.2 $\times$ 10$^{-4}$\\ 
\enddata
\tablecomments{
$^{a}$ 1 ML = 10$^{15}$ molecules cm$^{-2}$. We assume a 20\% error in the ice thickness (see Appendix \ref{sec:thickness_app}). 
$^{b}$ In the 2 keV electron irradiation experiments, this is the number of monolayers that absorb 95\% of the incident energy (Sect. \ref{sec:electron}). The electron penetration depth calculated for Exp. 1 was adopted throughout Experiments 2$-$7. 
}
\end{deluxetable*}{}

Table \ref{table_exp} summarizes the experimental simulations carried out for this paper. 
The irradiation of the H$_2$:CO:$^{15}$N$_2$ ice analogs took place in Experiments 1$-$3 (2 keV electrons), and 8 (Ly-$\alpha$ photons). 
%
%
Experiments 4$-$7 consisted in 2 keV electron irradiation of H$_2$:CO:N$_2$ ice mixtures 
using different combinations of molecular hydrogen (H$_2$ and D$_2$), carbon monoxide (CO, $^{13}$CO, and C$^{18}$O), and molecular nitrogen (N$_2$ and $^{15}$N$_2$) isotopologues, and were used to confirm the ice chemistry product assignments in the previous experiments. 
In addition, we also performed 2 keV electron and Ly-$\alpha$ photon irradiation of binary H$_2$:CO (Experiments 9 and 10, respectively), CO:$^{15}$N$_2$ (Experiments 11 and 12), and H$_2$:$^{15}$N$_2$ (Experiments 13 and 14) ice mixtures, to understand how the presence of each component affects the chemistry of the apolar ice layer. 

The experimental simulations were performed in the new SPACE TIGER\footnote{Surface Photoprocessing Apparatus Creating Experiments To Investigate Grain Energetic Reactions} setup with a 500 mm diameter stainless steel, ultra-high-vacuum (UHV) chamber custom made by  Pfeiffer Vacuum. 
The base pressure of $\sim$2 $\times$ 10$^{-10}$ Torr at room temperature is reached by 
a combination of magnetically levitated and scroll pumps. 
A detailed description of this setup can be found in \citet{pavlo20}, and the relevant features used for this work are described below.

\subsection{Ice sample preparation}\label{sec:ice}\label{sec:comp}

The ice samples were grown on a 12 mm diameter copper substrate  
(except for Experiments 1, 4, and 13, where a 10 mm diameter silver substrate was used)
at a temperature of 4.3 K (achieved by means of a closed-cycle He cryostat, Model DE210b-g, Advance Research Systems, Inc.), 
%
%
%
by exposing the substrate  to 
a gas mixture prepared in a independently pumped stainless steel mixing volume. 
%
The gas mixtures were introduced in the UHV chamber through a 10 mm diameter dosing pipe in close proximity ($<$2 mm) to the substrate.  We thus assumed that the size of the deposited ice samples was 10 mm diameter. 
%
For a reproducible dosing rate, we employed a laser drilled 7x7, 4 $\mu$m diameter pinhole array (Lenox Laser, custom made) placed between the mixing volume and the UHV chamber.
The dosing rate was proportional to the number of holes in the array, and the pressure in the mixing volume (up to 20 Torr); and inverse proportional to the square root of the species molar masses.  
The dosing process was terminated by shutting off the valve that allowed the gas mixtures to go through this pinhole array, and evacuating the small remaining volume between the valve and the array.
The ice composition was estimated to be the same as the composition of the gas mixture, assuming a sticking coefficient value of 1 for all the species onto the surface at $\sim$4 K. 
The gas mixture composition was measured in the UHV chamber before the ice deposition using a quadrupole mass spectrometer (QMS, Sect. \ref{sec:tpd}).
The measured QMS signals were transformed into partial pressures with a conversion factor previously calculated for the pure gases (see Appendix \ref{sec:comp_app} for the H$_2$, CO, and $^{15}$N$_2$ examples).
The ice analogs in Experiments 1$-$3, and 8$-$14 were deposited from gas mixtures composed by H$_2$ (gas, $>$99.99\% purity, Aldrich), CO (gas, 99.95\%, Aldrich), and $^{15}$N$_2$ (gas, 98\%, Aldrich). 
For the isotopically labeled ices in Experiments 4$-$7, we also used D$_2$ (gas, 99.96\%, Aldrich), $^{13}$CO (gas, 99\% $^{13}$C, $<$5\% $^{18}$O, Aldrich), C$^{18}$O (gas, 99\% $^{12}$C, 95\% $^{18}$O, Aldrich), and N$_2$ (gas, 99.99\%, Aldrich). 

\subsection{Electron irradiation of the ice samples}\label{sec:electron}

The deposited ice samples were electron irradiated in Experiments 1$-$7, 9, 11, and 13 using a ELG-2/EGPS-1022 low energy electron source system (Kimball Physics) with an average electron beam current of $\sim$54 nA (relative uncertainty of 10\%). 
The electron beam had an incidence angle of 56$^\circ$,  
resulting in an elliptical spot at the sample position,  
with a minor axis the same size as the ice sample, and a major axis $\sim$35\% larger\footnote{The size of the electron and photon beam spots was estimated with a phosphorus screen located at the sample position.}. 
The sample irradiation time was 45 min, so  
the ice samples were processed with $\sim$6.5 $\times$ 10$^{14}$ electrons with a fixed energy of 2 keV, leading to a total incident energy of $\sim$1.3 $\times$ 10$^{18}$ eV. 

The length of the electron irradiation experiment was meant to match the energy fluence directly deposited by the cosmic rays into the interstellar ice mantles during the typical lifetime of a dense cloud ($\sim$ 10$^6$ yr). 
The total irradiated energy fluence of $\sim$1.7 $\times$ 10$^{18}$ eV cm$^{-2}$  is similar to the estimated energy fluence deposited by the cosmic rays into the interstellar ice mantles during $\sim$1.3 $\times$ 10$^6$ yr  \citep[see][and ref. therein]{jones11}. 

The 2 keV electron penetration depth in the ice samples 
was lower than the total ice thickness, and was calculated with the CASINO v2.42 code \citep{drouin07} in terms of number of monolayers (1 ML = 10$^{15}$ molecules cm$^{-2}$). 
For example, in a H$_2$:CO:$^{15}$N$_2$ ice analog with a 1.8:1.0:1.0 composition (Exp. 1, Table \ref{table_exp}), 
95\% of the irradiated energy is absorbed in the first 380 ML of the ice. 
This means that only a fraction of the ice should be considered chemically active, since the bottom monolayers are barely processed by the 2 keV electrons.  
The number of monolayers of the ice sample effectively processed by the 2 keV electrons is higher in ices with a lower density, such as the H$_2$:CO ice of Exp. 9 (420 ML); 
while 95\% of the incident energy is absorbed in the first 220 ML of the CO:$^{15}$N$_2$ ice in Exp. 11, with a higher ice density. 
The electron penetration depth is indicated in the fifth column of Table \ref{table_exp}. 
The number of monolayers that were chemically active was taken into account when estimating conversion yields (Sect. \ref{sec:conv}).

\subsection{VUV photoprocessing of the ice samples}\label{sec:uv}

The ice analogs were VUV photon irradiated in Experiments 8, 10, 12, and 14.
Tunable VUV photons are generated in SPACE TIGER with a laser-based system \citep{pavlo20}. 
In particular, Ly-$\alpha$ photons were produced via two-photon resonance enhanced difference frequency mixing in krypton \citep{marangos90}. 
The number of irradiated Ly-$\alpha$ photons per second was measured with a Hamamatsu S10043 photodiode, and  
varied between 1.8$-$2.4 $\times$ 10$^{13}$ ph. s$^{-1}$ across the different experiments.  
The sample irradiation time in those experiments was set to 90$-$110 min. 
The ice samples were thus processed with $\sim$1.4 $\times$ 10$^{17}$ photons of $\sim$10.2 eV energy, leading to a total incident energy of $\sim$1.4 $\times$ 10$^{18}$ eV. 

As in the electron irradiation experiments, the length of the photon irradiation was set to match the energy fluence irradiated into the ice mantles  by the secondary UV field in dense cloud interiors. 
The laser beam 
had a normal incidence angle, and was $\sim$50\% the size of the ice sample surface, approximately. 
This resulted in a total incident energy fluence of $\sim$3.5 $\times$ 10$^{18}$ eV cm$^{-2}$ over the processed ice. 
For comparison, interstellar ices would be exposed to an energy fluence of $\sim$3.2 $\times$ 10$^{18}$ eV cm$^{-2}$ during a cloud lifetime of 10$^6$ years, considering a secondary UV flux in the interior of dense clouds of $\sim$10$^4$ photons cm$^{-2}$ s$^{-1}$ \citep{shen04}, and a photon energy of 10.2 eV.

The ice sample Ly-$\alpha$ photon absorbances in Experiments 8, 10, 12, and 14 were calculated according to the Beer-Lambert law:

\begin{equation}
    1 - \frac{I_t(Ly\alpha)}{I_0(Ly\alpha)} = 1 - e^{-\sigma_{abs}(Ly\alpha)N},
    \label{eqabs}
    \end{equation}{}

where $I_t(Ly$-${\alpha})$ and $I_0(Ly$-${\alpha})$ are the transmitted and incident Ly-$\alpha$ intensities, respectively; 
$\sigma_{abs}(Ly$-${\alpha})$  is the photon absorption cross-section at the Ly-$\alpha$ wavelength in cm$^2$; 
and $N$ is the ice thickness in terms of molecules cm$^{-2}$ (see Table \ref{table_exp}). 
$\sigma_{abs}(Ly$-${\alpha})$ 
is expected to be negligible for solid H$_2$ \citep{chuang18}, and for $^{15}$N$_2$ ices \citep[$\le$1.5 $\times$ 10$^{-21}$ cm$^2$,][]{gustavo14b}; 
while the value for CO ices is 1.0 $\times$ 10$^{-19}$ cm$^2$ \citep{gustavo14a}. 
The calculated Ly-$\alpha$ photon absorbance in Experiments 8, 10, and 12 was, approximately, 5\%, 6\%, and 7\%, respectively, while the absorbance in Exp. 14 was negligible. The absorbed energy in the Ly-$\alpha$ photon irradiation experiments is indicated in the last column of Table \ref{table_exp}.

\subsection{IR ice spectroscopy}\label{sec:ir}

The ice samples were monitored during the experimental simulations through reflection-absorption infrared spectroscopy (RAIRS) using a Bruker 70v Fourier transform infrared (FTIR) spectrometer and a liquid-nitrogen-cooled MCT detector. 
The spectra were averaged over 128 interferograms, and collected with a resolution of 1 cm$^{-1}$ in the 5000$-$800 cm$^{-1}$ range after deposition of the ice samples, every 5 min during electron processing, every 10$-$20 min during VUV photon processing, and every 5 min during warm-up of the processed ices (Sect. \ref{sec:tpd}). 

The IR beam 
had an incidence angle of 20$^\circ$ with respect to the substrate surface, resulting in an elliptical beam at the sample position, with a 14.6 mm major axis and a 5 mm minor axis.  
This means that 
the IR beam was smaller than the size of the ice sample surface and the 2 keV electron beam, but larger than the Ly-$\alpha$ photon beam.
%
The IR beam dilution of the ice chemistry products formed upon UV photon irradiation  
was taken into account 
for the quantification of 
the product conversion yields  
in Ly-$\alpha$ irradiated ices (Sect. \ref{sec:conv}).

\begin{table}[ht!]
\centering
\caption{Band strengths of selected features in pure ice IR spectra collected in transmission mode at 25 K, as reported in \citet{bouilloud15}. A  20\% uncertainty is assumed. 
\label{table_band}}
\begin{tabular}{ccc}
\hline
\hline
Molecule&Wavenumber&Band strength\\
&(cm$^{-1}$)&(cm molec$^{-1}$)\\
\hline
CO & 2139 & 1.1 $\times 10^{-17}$\\
$^{13}$CO & 2092 & 1.3 $\times 10^{-17}$\\
CO$_2$ & 2144 & 7.6 $\times 10^{-17}$\\
CH$_{4}$ &1304 & 8.0 $\times 10^{-18}$\\
H$_2$CO & 1720 & 1.6 $\times 10^{-17}$\\
\hline
\end{tabular}
\end{table}

\subsection{Conversion yields}\label{sec:conv}

To quantify the conversion of CO molecules into C-bearing products upon energetic processing of the H$_2$:CO:$^{15}$N$_2$, H$_2$:CO, and CO:$^{15}$N$_2$ ice samples, we 
calculated the ratio between the final product column density and the initial (active) CO column density ($N_f(X)$/$N_i(CO)$).
The column densities of the species with detected IR bands are proportional to the integrated IR absorbances:

\begin{equation}
N(X)=\frac{1}{A_X}\int_{band}{\tau_{\nu} \ d\nu},
\label{eqn}
\end{equation}

\noindent where $N(X)$ is the column density of the species $X$ in molecules cm$^{-2}$, $\tau_{\nu}$ is the optical depth of the absorption band (2.3 times the absorbance), and $A_X$ is the band strength of the IR feature in cm molecule$^{-1}$.

The band strengths $A_X$ of the IR features detected through RAIRS are different from those reported for the IR spectra collected in transmission mode  
%
but, similar to previous studies \citep[e.g.,][]{oberg09}, we assumed that the relative band strengths were the same in the IR spectra collected in transmission and reflection-absorption mode:

\begin{equation}
    \frac{N_f(X)}{N_i(CO)}=\frac{\int_X{\tau_{\nu} \ d\nu}}{\int_{CO}{\tau_{\nu} \ d\nu}}\frac{A_{CO}}{A_X}
    \label{eqy}
\end{equation}{}

Table \ref{table_band} presents the band strengths of selected IR features, with an assumed uncertainty of 20\% \citep{bouilloud15}. 

The IR absorbance in reflection mode shows a non-linear behavior with the species column densities above a certain threshold \citep{oberg09}. When this happens, the integrated absorbance is not proportional to the species column density, and Eq. \ref{eqn} is not valid. In our experimental simulations, this non-linear behavior prevented us from using the integrated absorbance of the CO IR feature ($\int_{CO}{\tau_{\nu} d\nu}$) in Eq. \ref{eqy} 
to determine the product conversion yields for Experiments 1$-$12. 
Instead, we used the small fraction of $^{13}$CO molecules deposited along with the high purity CO in every experiment to calculate $N_f(X)$/$N_i(^{13}CO)$, using the integrated absorbance of the $^{13}$CO IR feature at $\sim$2092 cm$^{-1}$ in Eq. \ref{eqy}. 
We subsequently multiplied this ratio by the $^{13}$CO/CO fraction of 1.4 $\times$ 10$^{-4}$,  
empirically determined using separate pure CO ice experiments where the IR spectra were collected in transmission mode. 
We assumed a 30\% uncertainty in the calculated $^{13}$CO/CO fraction, due to the 20\% uncertainty in the CO and $^{13}$CO IR band strengths \citep{bouilloud15}.

In addition to the systematic 20\% uncertainty in the band strength of the product IR features, and 30\% uncertainty in the calculated $^{13}$CO/CO fraction, we also considered a 25\% experimental uncertainty that accounted for the differences found in the conversion yields of experiments that should present the same results (Experiments 1$-$3, Appendix  \ref{sec:quan_app_qms}). 
Several sources of errors contributed to this experimental uncertainty, such as the uncertainty in the integrated $^{13}$CO IR absorbances (due to the low signal-to-noise ratio of the feature), as well as day-to-day changes in the precise experimental conditions. 
Altogether, we estimated a total uncertainty of 45\% in the reported product conversion yield values.

In addition, we also calculated the product ratios 
with respect to 
CO$_2$ 
($N_f(X)$/$N_f(CO_2)$), and compared the product branching ratios in different experiments.  

As for the product conversion yields, we  considered the 20\% systematic uncertainty in the band strengths of the product IR features, and the additonal 25\% experimental uncertainty, leading to a total uncertainty of 35\% in the reported product branching ratios. 
Only the additional experimental uncertainty needed to be taken into account when comparing conversion yields and product branching ratios of the same species across different experiments.
%

\subsection{Temperature Programmed Desorption of the processed ice samples}\label{sec:tpd}

After the energetic processing of the ice analogs, the samples were warmed from 4 K up to 200 K using a 50 W  silicon nitride cartridge heater rod (Bach Resistor Ceramics). 
A 2 K min$^{-1}$ heating rate was applied until the complete sublimation of the ice samples was achieved. 
The temperature of the ice samples was monitored by a LakeShore 336 temperature controller using a calibrated Si diode sensor  
with a 2 K estimated accuracy and a 0.1 K relative uncertainty.
The desorbing molecules were detected with a QMG 220M1 QMS (Pfeiffer, mass range 1$–$100 amu, resolution of 0.5 amu, located at  $\sim$13 cm from the substrate) leading to a Temperature Programmed Desorption (TPD) curve for each species. 
This allowed us to detect additional ice chemistry products whose IR features could not be unambiguously detected.
The initial ice components were monitored through their main mass fragments ($m/z$ = 2, H$_2$; $m/z$ = 4, D$_2$; $m/z$ = 28, CO and N$_2$; $m/z$ = 29, $^{13}$CO; and $m/z$ = 30,  C$^{18}$O and  $^{15}$N$_2$). 
The ice chemistry products were also monitored through their main mass fragment, but we note that in the case of H$_2$CO and H$_2$C$_2$O, the main mass fragment was not the molecular ion (Sect. \ref{sec:h2_co_n2}).  
 

In addition, we used the area under the collected CO TPD curve ($A_{TPD}(m/z = 28)$) to estimate the initial amount of CO molecules in Experiments 1$-$12, measured as the CO ice column density ($N(CO)$) in molecules cm$^{-2}$. 
The correspondence between $A_{TPD}(m/z = 28)$ and $N(CO)$ was previously calibrated (Appendix \ref{sec:thickness_app}). 
The initial H$_2$ and $^{15}$N$_2$ ice column densities were estimated from $N(CO)$ assuming that the initial ice composition was the same as the composition of the gas mixture used for ice deposition. 
In Experiments 13 and 14 the $^{15}$N$_2$ column density was calculated from the area $A_{TPD}(m/z = 30)$ under the corresponding TPD curve (Appendix \ref{sec:thickness_app}). 


\section{Results}\label{sec:results}

\subsection{Ice chemistry product identification in analogs of the apolar ice layer}\label{sec:h2_co_n2}

\begin{figure*}
    \centering
    \includegraphics[width=12cm]{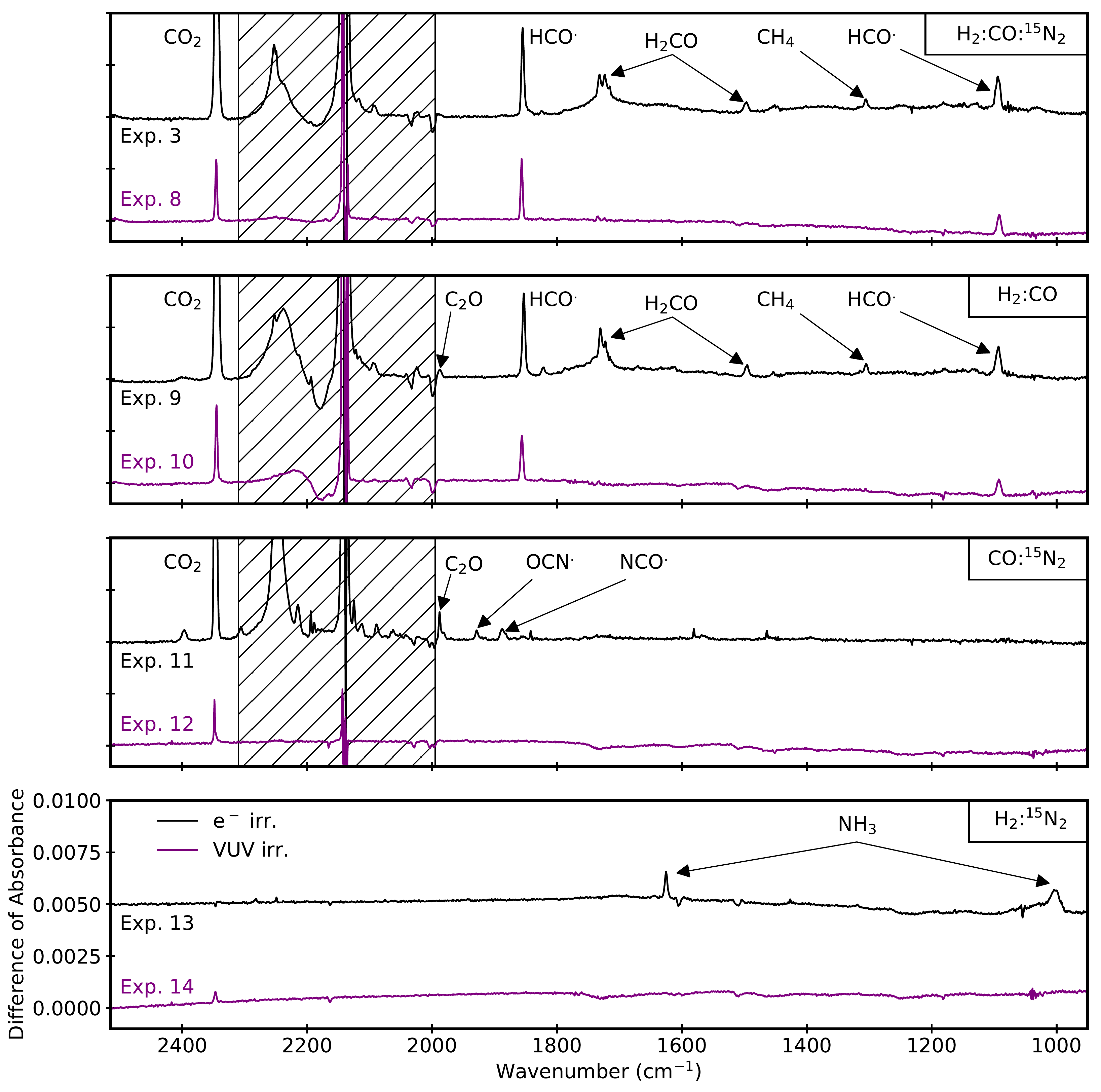}
    \caption{IR difference spectra obtained upon irradiation of a similar incident energy with 2 keV electrons (black) and Ly-$\alpha$ photons (purple) of H$_2$:CO:$^{15}$N$_2$, H$_2$:CO, CO:$^{15}$N$_2$, and H$_2$:$^{15}$N$_2$ ice samples (from top to bottom). 
    The IR spectra are offset for clarity. 
    IR band assignments are indicated in the panels.  
    The hatched region corresponds to the region of the spectrum dominated by CO and carbon chain oxide features (see Appendix  \ref{sec:co_app}).}
    \label{fig:ir}
\end{figure*}{}

The black curve in the top panel of Fig. \ref{fig:ir} shows the IR difference spectrum in the 2500$-$950 cm$^{-1}$ range upon $\sim$1.3 $\times$ 10$^{18}$ eV irradiation with 2 keV electrons of the H$_2$:CO:$^{15}$N$_2$ ice analog in Exp. 3 
(the results were similar in Experiments 1$-$3, Appendix \ref{sec:quan_app_qms}). 
The IR spectra in the 4000$-$950 cm$^{-1}$ range collected before and after irradiation are shown in Appendix \ref{sec:full_spec_app}.
The formation of CO$_2$, CH$_4$, and H$_2$CO at 4.3 K is evidenced by the growth of IR features at 2345 cm$^{-1}$, 1305 cm$^{-1}$, and 1725 cm$^{-1}$ and 1495 cm$^{-1}$, respectively. 
Two features due to the presence of unreacted HCO$^.$ radicals are also observed at 1855 cm$^{-1}$ and 1094 cm$^{-1}$. 
Additional IR features are detected in the 2310$-$1995 cm$^{-1}$ region of the spectrum, corresponding to carbon chain oxides. 
The formation of these species upon energetic processing of CO-bearing ice samples (including pure CO ices as shown in Appendix \ref{sec:co_app}) has been studied in \citet{sicilia12}.  
Hereafter we only discuss the formation of C$_2$O (see below) as a representative of the carbon chain oxide chemistry. 

\begin{figure*}
    \centering
    \includegraphics[width=12cm]{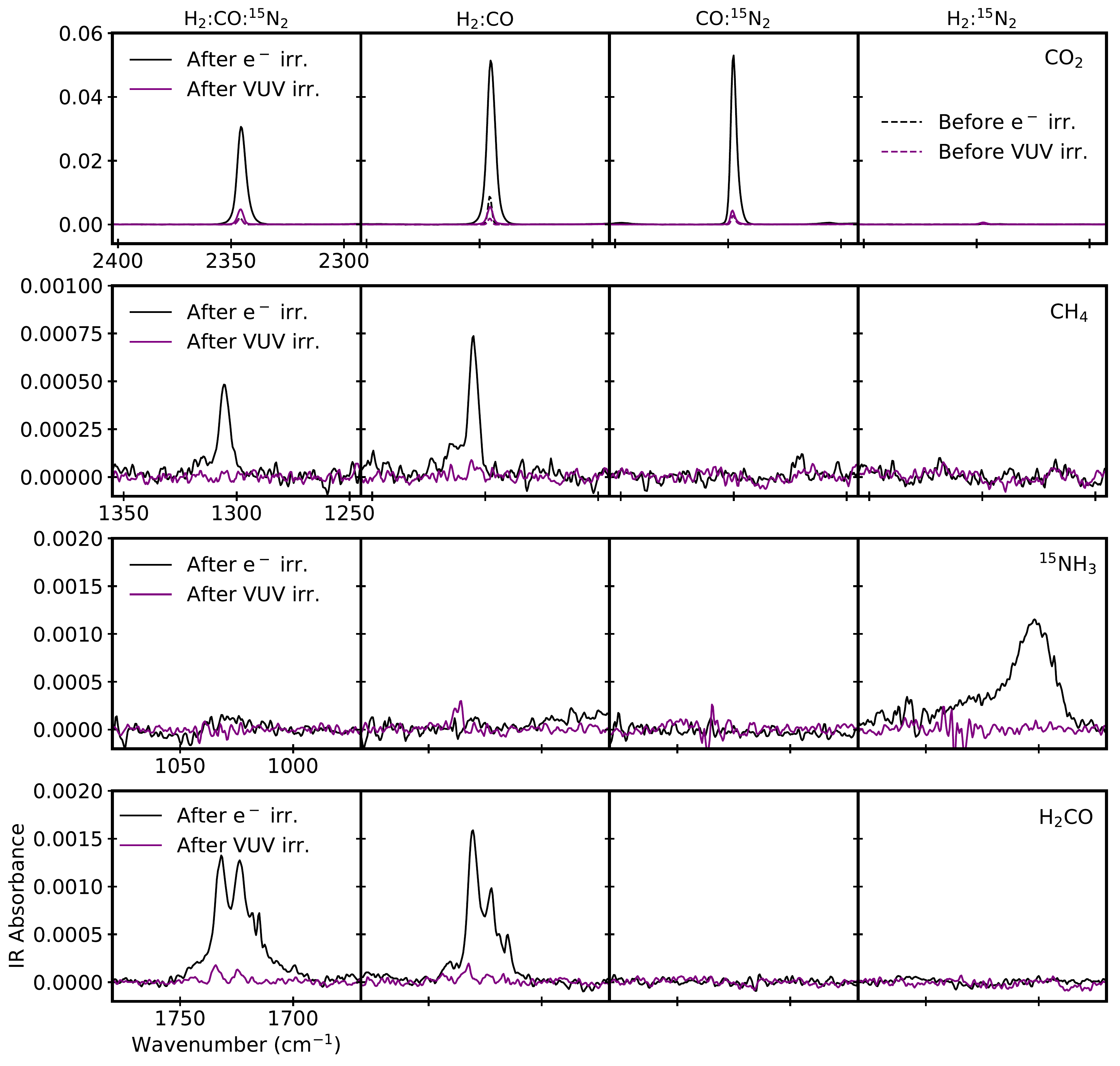}
    \caption{IR features of the CO$_2$, CH$_4$, $^{15}$NH$_3$, and H$_2$CO molecules (from top to bottom) formed upon irradiation of a similar incident energy with 2 keV electrons (solid black) and Ly-$\alpha$ photons (solid purple) of H$_2$:CO:$^{15}$N$_2$ (Experiments 3 and 8), H$_2$:CO (Experiments 9 and 10), CO:$^{15}$N$_2$ (Experiments 11 and 12), and H$_2$:$^{15}$N$_2$ (Experiments 13 and 14) ice samples (from left to right). The IR spectrum before the 2 keV electron (dashed black) and Ly-$\alpha$ photon (dashed purple) irradiation is included in the top panels, and shows the small amount of CO$_2$ contamination present in the ice sample (Appendix \ref{sec:co_app}). The IR spectra before and after 2 keV and Ly-$\alpha$ irradiation are nearly indentical in the top right panel, since no CO$_2$ formation was expected in a H$_2$:$^{15}$N$_2$}
    \label{fig:ir_bi}
\end{figure*}{}

An inset of the CO$_2$, CH$_4$, and H$_2$CO IR bands 
is shown in black in the left panels of Fig. \ref{fig:ir_bi}. 
A local baseline around each IR feature was subtracted using a \textit{spline} function with the IDL software. 
%
We note that a small amount of CO$_2$ contamination was present in the ice sample before irradiation, 
representing $\sim$0.01\% of the initial CO (see Appendix \ref{sec:co_app}).  
%
The double-peak structure of the H$_2$CO C=O stretching IR band at 1725 cm$^{-1}$ could be due to the apolar ice matrix in which the produced H$_2$CO was embedded 
since it was not observed for H$_2$CO molecules embedded in a H$_2$O polar ice matrix \citep{noble12}. 
In our experiments, the double peak structure was not observed at high temperatures, after desorption of the majority of the H$_2$, CO, and $^{15}$N$_2$ molecules (see Fig. \ref{fig:ir_det} in Appendix \ref{sec:products_app}). 

\begin{figure*}
    \centering
    \includegraphics[width=12cm]{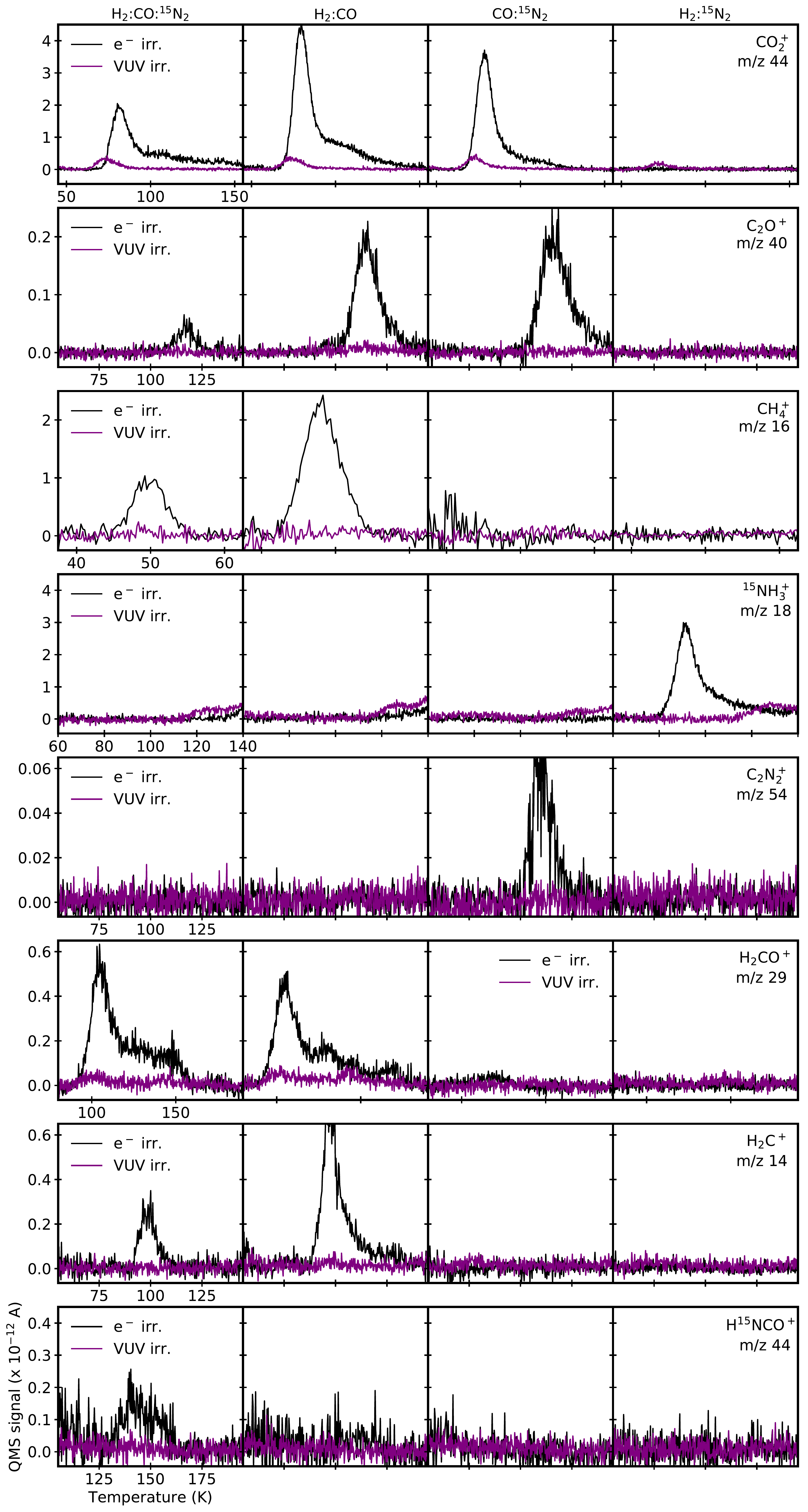}
    \caption{TPD curves of the CO$_2$, C$_2$O, CH$_4$, NH$_3$, C$_2$N$_2$, H$_2$CO, H$_2$C$_2$O, and H$^{15}$NCO  main mass fragments (from top to bottom, $m/z$ = 44, $m/z$ = 40, $m/z$ = 16, $m/z$ = 18, $m/z$ = 54, $m/z$ = 29, $m/z$ = 14, and $m/z$ = 44, respectively) during warm-up of H$_2$:CO:$^{15}$N$_2$ (Experiments 3 and 8), H$_2$:CO (Experiments 9 and 10), CO:$^{15}$N$_2$ (Experiments 11 and 12), and H$_2$:$^{15}$N$_2$ (Experiments 13 and 14) ice samples (from left to right) after the irradiation of a similar incident energy with 2 keV electrons (black) and Ly-$\alpha$ photons (purple).}
    \label{fig:tpd_bi}
\end{figure*}

The TPD curves of CO$_2$, CH$_4$, and H$_2$CO are shown in
Fig. \ref{fig:tpd_bi}, and were also baseline corrected using a \textit{spline} function with the IDL software. 
H$_2$CO presented two desorption peaks: one at 105 K, slightly below the multilayer desorption peak temperature measured in \citet{noble12}, and a second broad desorption peak above 130 K. We speculate that the second desorption peak could be due to the formation  of H$_2$CO dimers at temperatures above 35 K (see also Appendix \ref{sec:products_app}), although contribution from the fragmentation of desorbing H$^{15}$NCO molecules at $\sim$150 K (see below) could not be ruled out. 
%

{}

\begin{figure*}
    \centering
    \includegraphics[width=15cm]{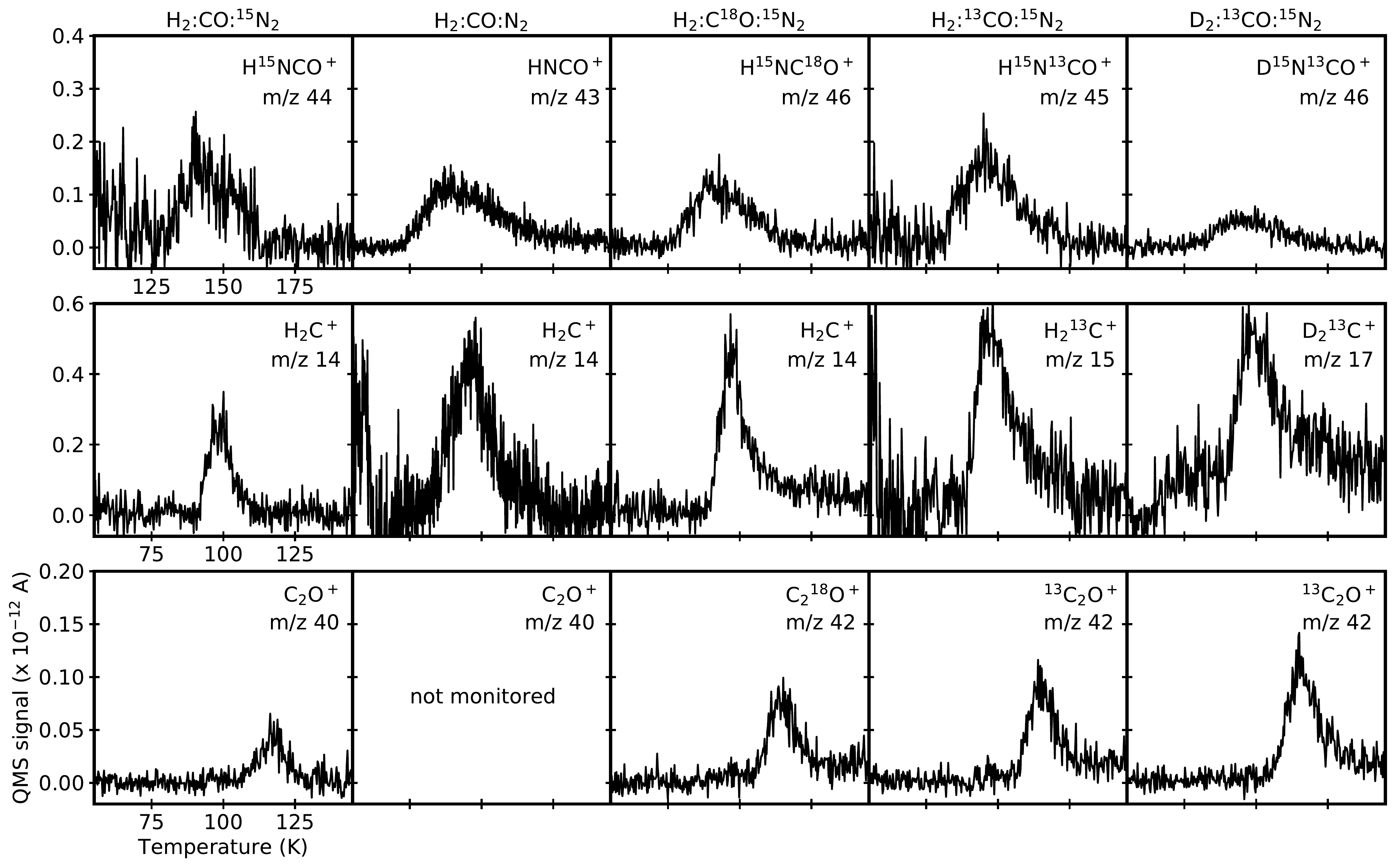}
    \caption{TPD curves of the H$^{15}$NCO, H$_2$C$_2$O, and C$_2$O main mass fragments (from top to bottom, respectively) during warm-up of different isotopically labeled three-component ice samples, from left to right:
    H$_2$:CO:$^{15}$N$_2$ (Exp. 3),H$_2$:CO:N$_2$ (Exp. 4),  H$_2$:C$^{18}$O:$^{15}$N$_2$ (Exp. 5), H$_2$:$^{13}$CO:$^{15}$N$_2$ (Exp. 6), and D$_2$:$^{13}$CO:$^{15}$N$_2$ (Exp. 7), after 2 keV electron irradiation. 
    The $m/z$ = 40 mass fragment was not monitored during warm-up of the electron irradiated H$_2$:CO:N$_2$ ice sample.}
    \label{fig:hnco_det}
\end{figure*}

Three additional products were detected using the QMS during the TPD of the processed ice analogs in Experiments 1$-$3: C$_2$O\footnote{No other carbon chain oxide molecules were monitored with the QMS}, H$_2$C$_2$O, and H$^{15}$NCO. 
The TPD curves of these product main mass fragments ($m/z$ = 40, C$_2$O; $m/z$ = 14, H$_2$C$_2$O; and $m/z$ = 44, H$^{15}$NCO\footnote{We assign the thermal desorption detected for the $m/z$ = 44 mass fragment at T $>$ 100 K to H$^{15}$NCO because it is the most stable isomer among HOC$^{15}$N, H$^{15}$NCO, HC$^{15}$NO, and HO$^{15}$NC. In previous experimental simulations where HNCO ice molecules were produced, the metastable isomers were only tentatively detected, if at all \citep[see, e.g.,][]{antonio14}. In addition, gas-phase observations of different astrophysical environments reveal an abundance of the metastbale isomers with respect to HNCO of less than 1\% \citep[and ref. therein]{quan10}.}) in Exp. 3 are also shown in
Fig. \ref{fig:tpd_bi}. 
The C$_2$O IR feature at 1989 cm$^{-1}$ \citep{palumbo08} was only tentatively detected at the edge of the hatched region in the top panel of Fig. \ref{fig:ir}. 
The H$_2$C$_2$O and H$^{15}$NCO IR features at 2129 cm$^{-1}$ and 2260 cm$^{-1}$, respectively \citep{broekhuizen04,hudson20}, overlapped with different CO and carbon chain oxide IR features in the hatched region, and could not be unambiguously detected. 
In order to confirm the assignments of the $m/z$ = 40, $m/z$ = 14, and $m/z$ = 44 desorptions in Fig. \ref{fig:tpd_bi},  
to C$_2$O, H$_2$C$_2$O, and H$^{15}$NCO (the thermal desorption detected at T $>$ 100 K), respectively, 
we energetically processed four additional ice analogs composed by different combinations of isotopically labeled H$_2$, CO, and N$_2$ molecules with 2 keV electrons (Experiments 4$-$7, see Table \ref{table_exp}), and checked that the same thermal desorption was observed for the expected  mass fragments, considering the shift in the mass of the main fragments according to their isotopic composition in each experiment.
The results are shown in Fig. \ref{fig:hnco_det}, confirming the formation of these species upon electron irradiation of H$_2$:CO:N$_2$ ice analogs. 

{}

\smallskip
After irradiation of similar incident energies with Ly-$\alpha$ photons (Exp. 7), two of the six ice chemistry products formed 
upon electron irradiation were detected: CO$_2$ and H$_2$CO.  
The CO$_2$ IR feature, as well as the two features due to unreacted HCO$^.$ radicals, are observed in the corresponding IR difference spectrum in the top panel of Fig. \ref{fig:ir}.
The inset of the CO$_2$ IR band in the top left panel of Fig. \ref{fig:ir_bi} shows that the amount of CO$_2$ molecules formed upon Ly-$\alpha$ photon irradiation is lower than in the 2 keV electron irradiation experiment. The formation of a small amount of H$_2$CO molecules is also observed in Fig. \ref{fig:ir_bi}. 
This suggests that the photon-induced chemistry took place to a lower extent, possibly, due to the low photon absorption cross-section of the ice sample at the Ly-$\alpha$ wavelength (Sect. \ref{sec:uv}).

\smallskip
Figures \ref{fig:ir}, \ref{fig:ir_bi}, and \ref{fig:tpd_bi} also show the IR difference spectra, the inset of the detected IR features, and the TPD curves of the ice chemistry products formed upon energetic processing of the binary ice mixtures in Experiments 9$-$14. 
The energetic processing of H$_2$:CO ice samples 
led to the formation of the same species as in the H$_2$:CO:$^{15}$N$_2$ ice analogs, except for H$^{15}$NCO. We note that 
the C$_2$O IR feature at 1989 cm$^{-1}$ was observed in the corresponding IR difference spectrum (Fig. \ref{fig:ir}), but it was blended with a CO IR band (Appendix \ref{sec:co_app}). 

Due to the lack of H$_2$ in the 
CO:$^{15}$N$_2$ ice samples, the H-bearing species CH$_4$, 
H$_2$CO, H$_2$C$_2$O, and  H$^{15}$NCO
could not be formed in Experiments 11 and 12. 
The IR difference spectrum after $\sim$1.3 $\times$ 10$^{18}$ eV irradiation with 2 keV electrons shows the formation of CO$_2$ and C$_2$O, as well as the unreacted radicals OCN$^.$ and NCO$^.$ (Fig. \ref{fig:ir}). 
In addition, thermal desorption  of C$_2$$^{15}$N$_2$ was also detected (Fig. \ref{fig:tpd_bi}, bottom panels), while  
only CO$_2$ formation was observed after irradiation with Ly-$\alpha$ photons.

In H$_2$:$^{15}$N$_2$ ices, $^{15}$NH$_3$ formation was observed after 2 keV electron irradiation in Exp. 13,  by means of both IR spectroscopy (Fig. \ref{fig:ir_bi}, right bottom panel),  and mass spectrometry during the subsequent TPD (Fig. \ref{fig:tpd_bi}). 
No other product was detected in this experiment (we note that the small amount of CO$_2$ observed in the right panel of Fig. \ref{fig:ir} was due to contamination). 

\subsection{Conversion yields and product branching ratios in 2 keV electron and Ly-$\alpha$ photon irradiated ices}\label{sec:co_conv}

\begin{figure*}
    \centering
    \includegraphics[width=11cm]{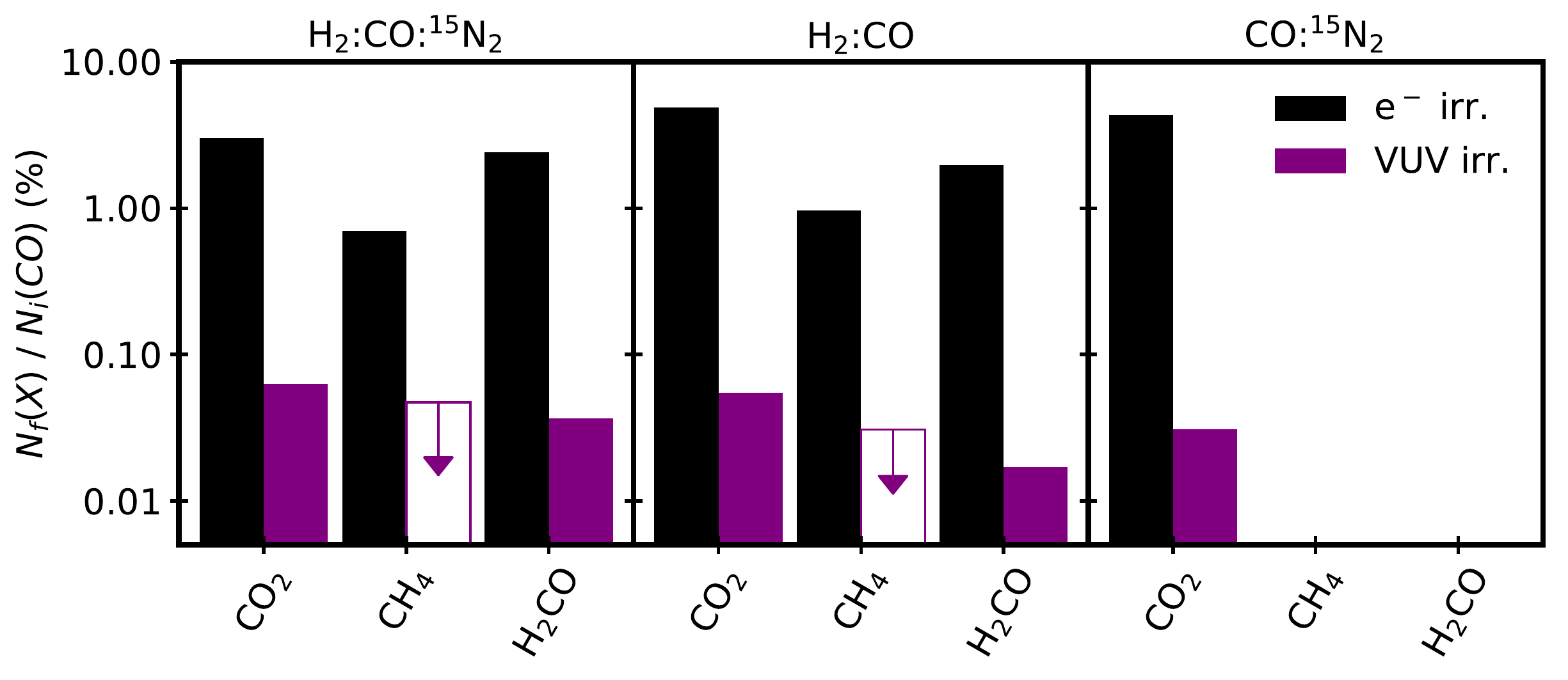}
    \caption{Product percent yields of the produced CO$_2$, CH$_4$, and H$_2$CO with respect to the initial CO ice column density after irradiation of similar incident energies with 2 keV electrons (black) and Ly-$\alpha$ photons (purple) of the different CO-bearing ice samples, from left to right: H$_2$:CO:$^{15}$N$_2$ (Experiments 3 and 8), H$_2$:CO (Experiments 9 and 10), and CO:$^{15}$N$_2$ (Experiments 11 and 12). 
    The percent yields were calculated from the integrated IR absorbances, assuming a 45\% total uncertainty (Sect. \ref{sec:conv}). 
    Empty bars indicate upper limits, and the arrows show the expected CH$_4$ formation in Experiments 8 and 10, assuming the same formation ratio with respect to CO$_2$ as in Experiments 3 and 9, respectively.
    }
    \label{fig:production_IR}
\end{figure*}{}

\begin{deluxetable}{cccccc}
\tablecaption{Product percent yields with respect to the initial CO ice column density\label{table_yields}}
\tablehead{
\colhead{Exp.} & \colhead{Ice comp.} & \colhead{Irradiation} & \multicolumn{3}{c}{Percent yield (\%)}\\
& & & \colhead{CO$_2$} & \colhead{CH$_4$} & \colhead{H$_2$CO}
}
\startdata
3 & H$_2$:CO:$^{15}$N$_2$ & 2 keV e$^-$ & 3.0 & 0.7 & 2.4 \\
8 & H$_2$:CO:$^{15}$N$_2$ & Ly-$\alpha$ & 0.063 & $<$0.05 & 0.035\\
\hline
9 & H$_2$:CO & 2 keV e$^-$ & 4.9 & 1.0 & 2.0\\
10 & H$_2$:CO & Ly-$\alpha$ & 0.055 & $<$0.03 & 0.017 \\
\hline
11 & CO:$^{15}$N$_2$ & 2 keV e$^-$ & 4.3 & \nodata & \nodata \\
12 & CO:$^{15}$N$_2$ & Ly-$\alpha$ & 0.031 & \nodata & \nodata \\
\enddata
\tablecomments{We assume a 45\% total uncertainty in the measured percent yields (Sect. \ref{sec:conv}).}
\end{deluxetable}{}

As described in Sect. \ref{sec:conv}, we used the conversion yield ($N_f(X)$/$N_i(CO)$) to quantify the conversion of CO molecules into products. 
In order to get the conversion yields, the measured $N_f(X)$/$N_i(^{13}CO)$ ratios were multiplied by the $^{13}$CO/CO fraction of $\sim$1.4 $\times$ 10$^{-4}$. 
The CO$_2$, CH$_4$, and H$_2$CO IR bands in Fig. \ref{fig:ir_bi}, as well as the IR feature corresponding to the initial $^{13}$CO molecules, 
were numerically integrated  using the composite Simpson's rule (\textit{integrate.simps} in Python).  
The resulting ratio of integrated absorbances ($\int_X{\tau_{\nu} \ d\nu}$/$\int_{^{13}CO}{\tau_{\nu} \ d\nu}$\footnote{The contribution of the initial CO$_2$ contamination to the final CO$_2$ abundance was taken into account.}) was used in Eq. \ref{eqy} along with the IR band strengths listed in Table \ref{table_band}.  
%
The integrated absorbances of the ice chemistry products formed upon Ly-$\alpha$ photon irradiation were corrected for the IR beam dilution, as mentioned in Sect. \ref{sec:ir}. 
%
%
%
%
The $N_f(X)$/$N_i(CO)$ percent yields at the end of Experiments 3 and 8$-$12 are presented in Fig. \ref{fig:production_IR} and Table \ref{table_yields}. 
We note that the percent yields in Experiments 3, 9, and 11 are calculated with respect to the fraction of the ice that is chemically active, i.e., the fraction of the ice that absorbs most (95\%) of the incident energy from the 2 keV electrons. 
This fraction is $\sim$15\% in Exp. 3, $\sim$20\% in Exp. 9, and $\sim$10\% in Exp. 11 (Sect. \ref{sec:electron}). 
The estimated percent yields should be considered as average percent yields in the processed ice. 

After $\sim$1.3 $\times$ 10$^{18}$ eV irradiation with 2 keV electrons of the H$_2$:CO:$^{15}$N$_2$ ice analog in Exp. 3, 3.0\%, 0.7\%, and 2.4\% of the initial CO molecules 
were converted into CO$_2$, CH$_4$, and H$_2$CO,  respectively. 
The total CO-to-product percent yield could be up to $\sim$10\% in electron irradiated ices, considering that the conversion yields of H$^{15}$NCO, H$_2$C$_2$O, C$_2$O, and the rest of carbon chain oxides were not calculated from the IR measurements.
The CO$_2$ and H$_2$CO percent yields upon Ly-$\alpha$ photon irradiation of the same ice sample in Exp. 8 were 50$-$60 times lower  
after the same incident energy. 
We note that only $\sim$5\% of the energy irradiated by the Ly-$\alpha$ photons was absorbed (Sect. \ref{sec:uv}). Therefore, even if the the formation of products upon Ly-$\alpha$ photon irradiation was as effective as in the 2 keV electron irradiation experiment, the expected conversion yields would be $\sim$20 times lower. 

\begin{figure}
    \centering
    \includegraphics[width=8.25cm]{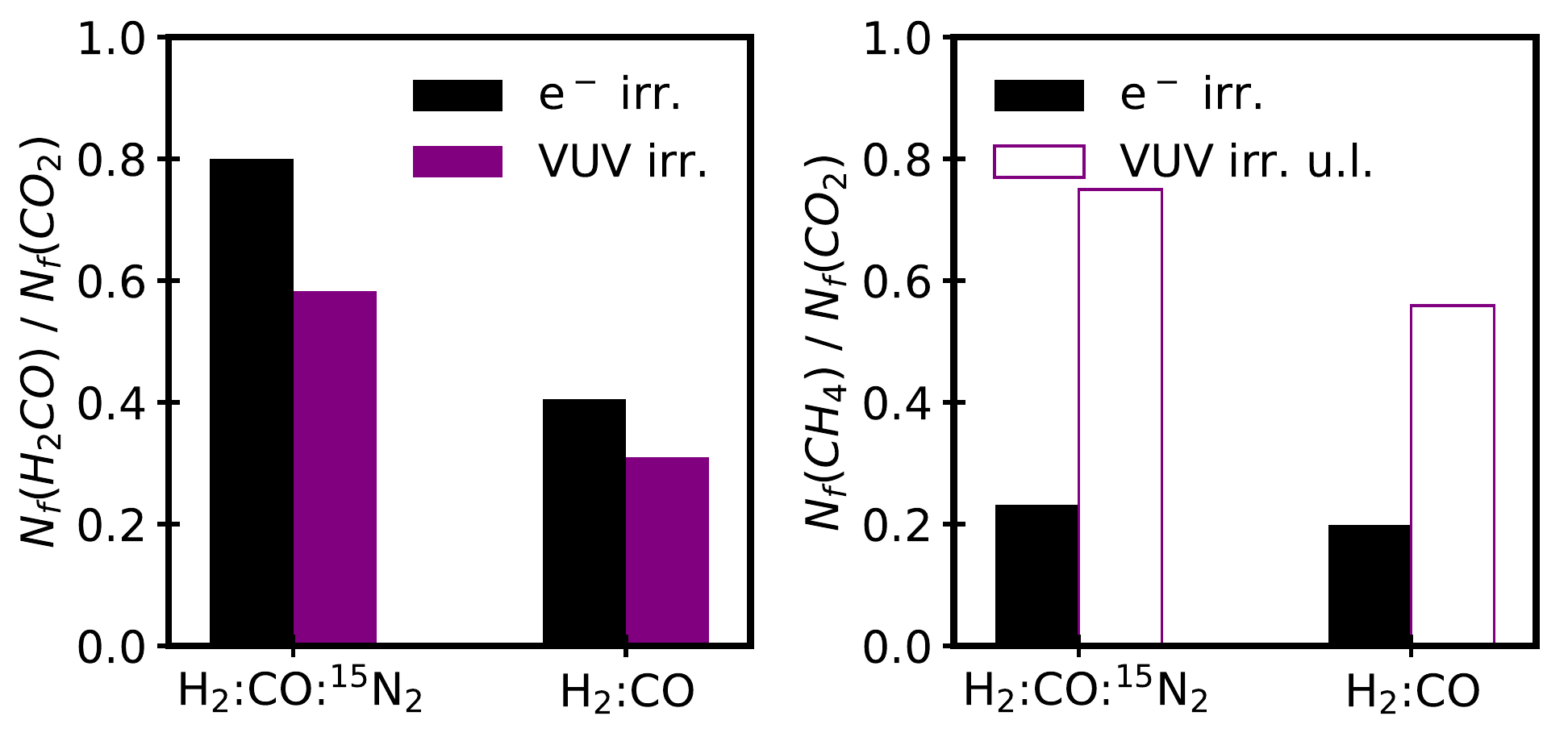}
    \caption{Product branching ratios of H$_2$CO (left panel) and CH$_4$ (right panel) with respect to CO$_2$ upon electron (black) and VUV (purple) irradiation of H$_2$:CO:$^{15}$N$_2$ ice samples in Experiments 3 and 8, and  H$_2$:CO ice samples in Experiments 9 and 10. A 35\% percent total uncertainty was assumed (Sect. \ref{sec:conv}).
    Empty bars indicate upper limits.}
    \label{fig:production_h2co}
\end{figure}{}

While the CO$_2$ and H$_2$CO conversion yields in the Ly-$\alpha$ photon irradiated ice sample were 1$-$2 orders of magnitude lower than in the 2 keV electron irradiated ice, the $N_f(H_2CO)$/$N_f(CO_2)$ product branching ratio was quite similar. The difference in the relative formation of H$_2$CO with respect to CO$_2$ was within the 25\% experimental uncertainty (Fig.  \ref{fig:production_h2co}, left panel). 
%
The non-detection of CH$_4$ after Ly-$\alpha$ photon irradiation of the ice samples was not informative, since the reported upper limits (empty bars in Figures \ref{fig:production_IR} and \ref{fig:production_h2co}) were above the expected yields if the same product branching ratio with respect to CO$_2$ as in the corresponding 2 keV irradiation experiments was assumed.

\smallskip

The CO$_2$, CH$_4$, and H$_2$CO percent yields in the 2 keV electron irradiated H$_2$:CO ice (Exp. 9) 
were 4.9\%, 1.0\%, and 2.0\%, respectively.  
These percent yields 
were 
two orders of magnitude lower for CO$_2$ and H$_2$CO upon Ly-$\alpha$ photon irradiation of the same ice sample (Exp. 10). 
%
The relative formation of H$_2$CO with respect to CO$_2$ was similar in the 2 keV electron and Ly-$\alpha$ photon irradiated H$_2$:CO ices (Experiments 9 and 10, Fig. \ref{fig:production_h2co}, left panel). 
%
The CO$_2$ percent yield was 4.3\% in the 2 keV electron irradiated CO:$^{15}$N$_2$ ice (Exp. 11, similar to the previous experiments),  
and two orders of magnitude lower
(0.031\%) upon Ly-$\alpha$ photon irradiation.

\subsection{CO$_2$ formation kinetics}\label{sec:kinetics}

\begin{figure*}
    \centering
    \includegraphics[width=12cm]{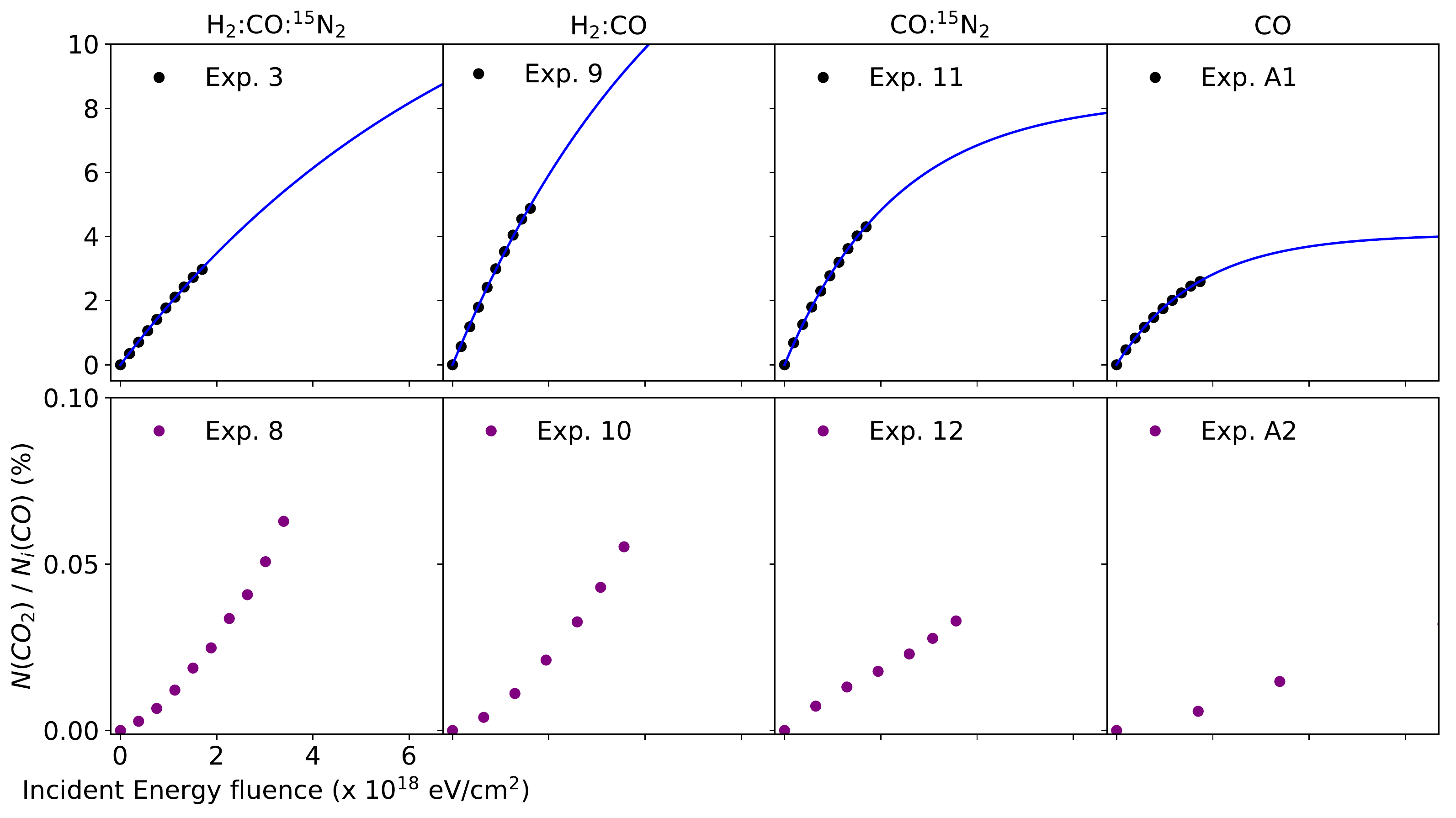}
    \caption{CO$_2$ growth curves in terms of the conversion yield versus the incident energy fluence upon electron (black, top panels) and VUV photon (purple, bottom panels) irradiation of H$_2$:CO:$^{15}$N$_2$, H$_2$:CO, CO:$^{15}$N$_2$, and pure CO ice samples (from left to right). 
    The Ly-$\alpha$ flux was not measured in Exp. A2, and we adopted the same value as in Exp. 7. 
    The best-fit pseudo-first order kinetic model is shown in the top panels (blue solid lines). 
    }
    \label{fig:kinetics}
\end{figure*}{}

\begin{deluxetable}{ccccccc}
\tablecaption{Summary of the best-fit CO$_2$ formation steady state percent yields and cross sections upon electron irradiation of different ice samples. \label{table_kin}}
\tablehead{
\colhead{Exp.} & \colhead{Ice comp.} & \colhead{$\frac{N_{ss}(CO_2)}{N(CO)}$ (\%)} & \colhead{$\sigma_{form}$ (cm$^2$)}}
\startdata
3 & H$_2$:CO:$^{15}$N$_2$ & 15 $\pm$ 3   & [1.3 $\pm$ 0.3] $\times$ 10$^{-19}$\\
\hline
9 & H$_2$:CO & 18 $\pm$ 3 & [2.0 $\pm$ 0.4] $\times$ 10$^{-19}$\\
\hline
11 & CO:$^{15}$N$_2$ & 8.3 $\pm$ 0.2 & [4.3 $\pm$ 0.2] $\times$ 10$^{-19}$\\
\hline
A1 & CO &  4.1 $\pm$ 0.1  & [5.9 $\pm$ 0.2] $\times$ 10$^{-19}$\\
\enddata
\tablecomments{Fit errors do not take into account the 45\% total uncertainty in the conversion yield values.}
\end{deluxetable}{}

In the case of CO$_2$, we had sufficient signal-to-noise ratio from the beginning to the end of the experimental simulations to study the formation kinetics across the different experiments, and use that to further explore the effects of the different energy sources and ice compositions.
Fig. \ref{fig:kinetics} shows the CO$_2$ growth curves in terms of the conversion yield 
versus the incident energy fluence 
in the CO-bearing ice samples (Experiments 3, and 8$-$12), 
as well as in 
pure CO ices (Experiments A1 and A2, Appendix \ref{sec:co_app}).

The CO$_2$ growth curves upon 2 keV electron irradiation of the studied ice samples (Fig. \ref{fig:kinetics}, top panels) could be parameterized with a pseudo-first order kinetic model \citep{ioppolo09}: 

\begin{equation}
\frac{N(CO_2)}{N_i(CO)} = \frac{N_{ss}(CO_2)}{N_i(CO)} \times (1-e^{-\sigma_{form} E})
\label{eqnh2cho}
\end{equation}

where $E$ is the incident energy fluence in eV cm$^{-2}$. $N_{ss}(CO_2)$/$N_i(CO)$ represents the CO$_2$ steady state conversion yield, i.e., the highest conversion yield that could be reached in an environment where the ices were exposed to higher energy fluences than the typical values expected in dense cloud interiors (Sect. \ref{sec:electron}). The parameter $\sigma_{form}$ is the apparent formation cross-section in cm$^2$. In a pseudo-first order kinetic model, this parameter indicates how fast the corresponding steady state conversion yield is reached, and cannot be used to determine the absolute formation rate. 
The best-fit steady state conversion yields and apparent formation cross sections are presented in Table \ref{table_kin}. 
%

The formation of CO$_2$ molecules in the 2 keV electron irradiated experiments seemed to be enhanced by the presence of H$_2$ molecules (Fig. \ref{fig:kinetics}, top panels). 
The steady state CO$_2$ percent yields in the H$_2$:CO:$^{15}$N$_2$ ice sample (Exp. 3) and the H$_2$:CO binnary mixture (Exp. 9) were more than a factor of 3 higher than the value in a pure CO ice (Exp. A1).  
The conversion from CO to CO$_2$ in the presence of H$_2$ molecules could proceed through the HO-CO intermediate, as it is
the case in other proposed CO$_2$ ice formation scenarios
\citep[for example, the CO + OH$^.$ reaction that is expected to take place in the H$_2$O-rich ice layer,][]{ioppolo11}. The formation of the HO-CO intermediate significantly
decreases the CO$_2$ formation activation barrier
compared to the CO + O reaction \citep{garrod11}, and could be the reason behind the observed enhancement in our experimental simulations. 

Even though the presence of $^{15}$N$_2$ molecules in the ice samples seemed to increase the formation of CO$_2$ when comparing the steady state percent yield of the CO:$^{15}$N$_2$ binary ice mixture (Exp. 11) with that of a pure CO ice (Exp. A1), the difference in the steady state percent yields were not very significant considering the 25\% uncertainty found between experiments (Appendix \ref{sec:quan_app_qms}). In addition, this effect  was negligible in ice samples where H$_2$ molecules were also present (Experiments 3 and 9, Table \ref{table_kin}). 
More experiments would be needed to constrain the effect of the $^{15}$N$_2$ molecules on the CO$_2$ formation.

The formation of CO$_2$ molecules upon Ly-$\alpha$ photon irradiation also seemed to be enhanced in H$_2$-bearing ices when comparing the H$_2$:CO:$^{15}$N$_2$ ice analog and the H$_2$:CO binary mixture with a pure CO ice, and  the CO:$^{15}$N$_2$ binary mixture, while the presence of $^{15}$N$_2$ did not have any significant effect (Fig. \ref{fig:kinetics}, bottom panels). 
However, the same kinetic model could not be used for the CO$_2$ growth curves in Ly-$\alpha$ photon irradiation experiments.


\section{Discussion}\label{sec:disc}

\subsection{Electron and photon processing of the apolar ice layer}

Electrons and UV photon irradiation both resulted in chemical reactions in the apolar ice analogs. 
Based on the observed H$_2$CO product branching ratio with respect to CO$_2$, the induced chemistry was not significantly affected by the nature of the energetic processing (Sect. \ref{sec:co_conv}), 
even though, unlike the 2 keV electrons, the energy of the Ly-$\alpha$ photons ($\sim$10.2 eV) was not enough to directly dissociate the CO ice molecules
\citep[the CO dissociation energy is 11.09 eV, see][and ref. therein]{hector19}.
%
%
Previous works comparing the energetic processing of H$_2$O-rich polar ices by cosmic-ray analogs and UV photons had found a general agreement in the induced ice chemistry \citep[see, e.g.,][]{gerakines01}. However, a significant decrease in the measured conversion yield is reported in the literature for selected products in UV irradiated apolar ices after similar absorbed energies.   
For example, the formation of the carbon chain oxide C$_3$O$_2$ presented a 17 times lower yield upon UV photolysis of CO-bearing ices compared to proton irradiation of the same samples  \citep{gerakinesMoore01}, while the formation of N$_3^.$ radicals was not observed upon UV irradiation of N$_2$-rich ices, in contrast with radiolysis experiments of the same samples \citep{hudson02}. 
The differences in the measured yields could indeed be linked to the inability of the UV photons to break the intramolecular bonds in CO and N$_2$ molecules. 
In our case,
the $\sim$50$-$100 times lower conversion in Ly-$\alpha$ photon irradiated ices compared to the 2 keV electron irradiation experiments could to some extent be explained by the low photon absorption cross-section of the ices at this particular wavelength, since  
only $\sim$5\% (i.e., a 1/20 fraction) of the irradiated energy was absorbed by the CO molecules (Sect. \ref{sec:uv}). 
We speculate that broad-band UV photon irradiation could lead to conversion yields within a factor of a few compared to the electron irradiated ices, although this would need to be verified. 

The CO$_2$ growth curves in 2 keV electron irradiated ices could be parameterized with a pseudo-first order kinetic model, but this model did not fit the growth curves in Ly-$\alpha$ photon irradiation experiments (Sect. \ref{sec:kinetics}). 
However these apparent differences in the formation kinetics could be misleading, since different parts of the CO$_2$ growth curves were probed in each case. 
If the CO$_2$ percent yields in Fig. \ref{fig:kinetics} were presented as a function of the absorbed energy fluence instead of the irradiated fluence, the complete growth curves of the Ly-$\alpha$ photon irradiation experiments would roughly correspond to the 
first measured percent yield in the 2 keV electron irradiated ices. 
We can thus speculate that the CO$_2$ formation in Ly-$\alpha$ photon irradiation experiments could be in an earlier regime compared to the formation after similar irradiated energy fluences with 2 keV electrons, and it could eventually evolve to a pseudo-first order kinetic curve at higher energy fluences. 
Additional data using broad-band UV irradiation and/or low-flux electron irradiation is needed to check whether there are any kinematic differences between the two types of energetic processing. 
%

During the dense cloud lifetime,  
the energy directly transferred from the incoming cosmic rays into the electronic system of the ice molecules 
is thought to be similar to
the incident energy experienced by the ice mantles from the cosmic-ray induced secondary UV field \citep{moore01}. 
From the results presented in this paper, 
keV electrons produced by the interaction of cosmic rays with the ice molecules \citep{bennett05} 
are probably more important for the apolar ice layer processing than 
the secondary UV field induced by the interaction of the cosmic rays with the gas-phase H$_2$ molecules \citep{cecchi92,shen04}.

\subsection{Effect of the different components in the ice chemistry}\label{sec:branch}

Experiments with binary ice mixtures were performed to understand how the presence of the different components affected the ice chemistry. 
In the absence of H$_2$ molecules, C$_2$$^{15}$N$_2$ was detected following electron irradiation of the CO:$^{15}$N$_2$ ice sample (Fig. \ref{fig:tpd_bi}).   
In addition, $^{15}$NH$_3$ molecules were only observed in the H$_2$:$^{15}$N$_2$ processed ice, while  
neither was detected in the mixture with three components. 
This suggests that the  products of CO and H$_2$ irradiation preferably react with each other instead of with $^{15}$N$_2$ or its dissociation products, 
leading to a quenched formation of $^{15}$NH$_3$ and C$_2$$^{15}$N$_2$ in H$_2$:CO:$^{15}$N$_2$ ices. 
Formation of $^{15}$NH$_3$ in H$_2$:CO:$^{15}$N$_2$ ice samples, and subsequent reaction with H$^{15}$NCO leading to the formation of OCN$^-$ and NH$_4^+$ probably did not take place, since 
the NH$_4^+$ IR feature at 1485 cm$^{-1}$ was not observed in Exp. 3 (Fig. \ref{fig:ir}). 
The OCN$^-$ IR feature at 2166 cm$^{-1}$ overlaps with different CO and carbon chain oxides IR bands, and could not be unambiguously detected either.

The CH$_4$ and H$_2$CO product branching ratios with respect to CO$_2$ upon energetic processing of H$_2$:CO:$^{15}$N$_2$ ice analogs and H$_2$:CO binary mixtures were used to explore the effect that the  $^{15}$N$_2$ molecules had on the ice chemistry. 
The relative product formation of H$_2$CO
with respect to CO$_2$ could be decreased in the absence of $^{15}$N$_2$ molecules 
(Fig. \ref{fig:production_h2co}, left panel). 
However, 
we note that the difference found in the product branching ratios was just above the 25\% experimental uncertainty. 
The 
CH$_4$ product branching ratio 
was
the same, within the experimental uncertainties, 
upon 2 keV electron irradiation of H$_2$-bearing, CO ices with or without $^{15}$N$_2$. 
%
%

On the other hand, a higher carbon chain oxide production in the 2 keV electron irradiated H$_2$:CO and CO:$^{15}$N$_2$ binary ice mixtures compared to the H$_2$:CO:$^{15}$N$_2$ ice analog was evidenced by the stronger 
C$_2$O IR feature observed in
the corresponding IR difference spectra (at the edge of the hatched region in Fig. \ref{fig:ir}).  
This was confirmed by the higher QMS signal detected during thermal desorption of this species in the corresponding experiments (Fig. \ref{fig:tpd_bi}). 

\subsection{Alternative pathways to H-atom addition reactions for the CH$_4$, NH$_3$, and H$_2$CO ice formation}

CH$_4$ has been detected in interstellar ice mantles with abundances of $\sim$5\% with respect to H$_2$O ice \citep[and ref. therein]{boogert15}. 
The strong correlation between the observed CH$_4$ and H$_2$O ice column densities, along with the broad profile of the solid-phase CH$_4$ IR feature, suggest that this species is present in the H$_2$O-rich polar ice layer, and supports a scenario where CH$_4$ molecules are formed on the surface of dust grains through successive H-atom additions to C atoms during the formation of the H$_2$O-rich ice layer \citep{oberg08}. 
This formation pathway has been recently probed through experimental simulations in \citet{qasim20}. 
In this work, we show that CH$_4$ molecules are also formed upon energetic processing of H$_2$-bearing, CO ices. 
The produced CH$_4$ molecules represented less than 1\% of the initial CO molecules 
(Table \ref{table_yields}). 
Therefore, we do not expect this formation pathway to significantly contribute to the total formation of CH$_4$ ice, in line with the weak correlation of the observed CH$_4$ and CO ice column densities \citep{oberg08}. 

The observed NH$_3$ ice abundance with respect to H$_2$O ice molecules is $\sim$7\%. 
NH$_3$ is also believed to be present in the H$_2$O-rich polar ice layer after formation through H-atom hydrogenation of N atoms during the accretion of the H$_2$O-rich ice layer \citep{boogert15}. 
%
%
As explained above, $^{15}$NH$_3$ molecules are not formed 
in the presence of CO molecules, and we thus not expect the NH$_
3$ molecules to be present in the CO-rich ice layer. 
%
%
The formation through apolar ice energetic processing would only be expected if CO and N$_2$ ices were partially segregated, which may occur due to the slightly lower binding energy of N$_2$ molecules compared to CO molecules in pure and mixed CO:N$_2$ ices \citep{bisschop06}.  

H$_2$CO has been likely detected in interstellar ice mantles, with estimated abundances ranging from 2\% to 7\% with respect to H$_2$O ice in different environments \citep[and ref. therein]{boogert15}. 
However, current observations are not of sufficient quality to determine if H$_2$CO would be present in the H$_2$O-rich or the CO-rich ice layers \citep[and ref. therein]{qasim19c}. 
The most accepted H$_2$CO ice formation pathway is the H-atom hydrogenation of CO molecules \citep[see, e.g.,][]{watanabe02,fuchs09}. 
H$_2$CO is thus more likely to be found in the CO-rich apolar ice layer \citep{cuppen09}.  
%
More recently, \citet{chuang18} proposed the VUV photon processing of H$_2$:CO ices 
as an additional H$_2$CO ice formation pathway.    
%
In this work we show that 
H$_2$CO also forms upon electron irradiation of H$_2$-bearing, CO ices. The conversion with respect to the initial CO molecules in a H$_2$:CO:$^{15}$N$_2$ ice analog ($\sim$2\%, Table \ref{table_yields}) and  the  
product branching ratio with respect to CO$_2$\footnote{We note that approximately one third of the observed CO$_2$ ice in the interstellar medium is present in a CO environment \citep{pontoppidan08}, and could be formed upon cosmic ray irradiation of CO ices \citep{jamieson06}.}  (Fig. \ref{fig:production_h2co}, left panel) 
are enhanced compared to H$_2$:CO ices. 
More experimental simulations aiming to characterize this formation pathway are needed in order to properly address the relative contribution of the different H$_2$CO formation scenarios.

\subsection{H$_2$C$_2$O and HNCO formation in the interstellar medium}

H$_2$C$_2$O is thought to be chemically related to other organics such as CH$_3$CHO and CH$_3$CH$_2$OH, that belong to the family of O-bearing COMs described by the chemical formula C$_2$H$_n$O \citep{chuang20}. 
H$_2$C$_2$O has been observed in the gas-phase across different astrophysical environments. 
The higher H$_2$C$_2$O abundance observed in the cold envelopes around high-mass protostars compared to their hot cores \citep{ruiterkamp07}, along with its detection in translucent clouds, before the accretion of the CO-rich ice layer \citep{turner99}, hint at a solid-phase formation contemporary to the H$_2$O-rich ice layer formation, followed by subsequent desorption to the gas phase. 
One proposed formation pathway in line with this scenario is the reaction of simple hydrocarbons with OH$^.$ radicals, as recently presented in \citet{chuang20}.  
%
An alternative H$_2$C$_2$O formation scenario in the CO-rich apolar ice layer is presented in this work. 
However, the lack of a proper quantification of the conversion yield prevents us from addressing the contribution of this formation pathway to the solid-phase H$_2$C$_2$O formation.

The formation of HNCO in the ISM is of particular interest from a prebiotic chemistry perspective, since it contains the peptide bond (-(H-)N-C(=O)-) that links amino acids into proteins. 
The observed HNCO (and two of its metastable isomers, HOCN and HCNO) gas-phase abundances across different astrophysical environments cannot be explained with current gas-phase astrochemical models, and contribution from solid-phase formation pathways is needed to some degree depending on the environment \citep{tideswell10,quan10,quenard18}. 
HNCO has been observed in comets \citep{mumma11}, but in interstellar ices only the related species OCN$^-$ has been detected \citep{pontoppidan03,vanBroekhuizen05} with a typical abundance of $\sim$0.6\% with respect to H$_2$O \citep{boogert15}. 
\citet{broekhuizen04} proposed that solid-phase OCN$^-$ is formed from HNCO in the presence of NH$_3$ molecules, through an acid-base like reaction that would take place in the H$_2$O-rich polar ice layer. 
However, a possible correlation between CO and OCN$^-$ is reported in \citet{oberg11}, and strong enhancements of the OCN$^-$ ice abundance are observed in cold regions after accretion of the CO-rich apolar layer \citep{boogert15}, which suggests that OCN$^{-}$, and in extension HNCO formation is occurring in the apolar ice layer.
Based on laboratory experiments, 
the hydrogenation of N atoms,   
and the simultaneous hydrogenation and UV irradiation of NO molecules in CO-rich ice analogs have been proposed as possible formation pathways for HNCO in the apolar ice layer  \citep{fedoseev15b,fedoseev16}. 
However, it is not obvious that these pathways can fully account for the observed abundance of HNCO in the ISM. 
In this work we show that the electron irradiation of a H$_2$:CO:$^{15}$N$_2$ ice analog 
also results in H$^{15}$NCO formation.  
Even though the formation of other $^{15}$N-bearing species was quenched upon energetic processing of a H$_2$:CO:$^{15}$N$_2$ ice analog (Sect. \ref{sec:branch}), the formation of H$^{15}$NCO was confirmed during the TPD of the processed ice sample (Fig. \ref{fig:hnco_det}). However, the overlapping of the corresponding IR feature with different CO and and carbon chain oxide IR bands prevented us from calculating the conversion yield. 
A proper quantification of the HNCO conversion yield will help to properly estimate the contribution of this formation pathway to the budget of this species in the ISM.

\section{Conclusions}\label{sec:conc}

\begin{enumerate}
    \item Interstellar ice mantles are characterized by H$_2$O-rich (polar) and CO-rich (apolar) layers. Energetic processing of polar ices have long been known to induce a complex organic chemistry. In this study we show that a complex ice chemistry can also be induced by energetic processing of the 
    CO-rich apolar layer if H$_2$ molecules are present, 
    and can proceed in very cold environments (down to $\sim$4 K). 
    
    \item Several products were identified in our experiments simulating the electron processing of a H$_2$:CO:$^{15}$N$_2$ ice analog, including simple species as CO$_2$, C$_2$O (among other carbon chain oxides), and CH$_4$; and the more complex organics H$_2$CO (a COM precursor), H$_2$C$_2$O, and H$^{15}$NCO.  In particular, H$^{15}$NCO is an interesting molecule from a prebiotic chemistry perspective, since it contains the peptide bond ($-$(H$-$)N$-$C($=$O)$-$). 
    
    \item The CO conversion into these products in the 2 keV electron irradiation experiment accounts for $\sim$5$-$10\% of the initial CO exposed to the irradiation. 
    More experimental simulations aiming to quantitatively characterize these formation pathways are needed in order to asses their relative contribution to the interstellar budget of, especially, the organics H$_2$CO, 
    and H$^{15}$NCO. 
    
    \item The CO$_2$ and H$_2$CO formation is 1$-$2 orders of magnitude lower in Ly-$\alpha$ photon irradiated ices compared to 2 keV electron irradiated ices when the incident energy is similar, which may be due to the low UV-photon absorption cross-section of the ice at this wavelength; while  
    the relative branching ratio is similar in both cases. 
    From the results presented in this paper, keV electrons produced by the interaction of cosmic rays with the ice molecules probably contribute to the apolar ice layer chemistry to a larger extent than the cosmic-ray-induced secondary UV field.

    
    
    \item Formation of $^{15}$NH$_3$ and C$_2$$^{15}$N$_2$ is only observed in the absence of CO and H$_2$ molecules, respectively.  
    This suggests that chemical pathways involving both H$_2$ and CO  molecules are preferred over those involving only H$_2$ and $^{15}$N$_2$ or CO and $^{15}$N$_2$ molecules.

\end{enumerate}

\acknowledgments

This work was supported by an award from the Simons Foundation (SCOL \# 321183, KO).

\appendix

\section{Ice sample preparation}

\subsection{Ice composition}\label{sec:comp_app}

\begin{figure}
    \centering
    \includegraphics[width=11cm]{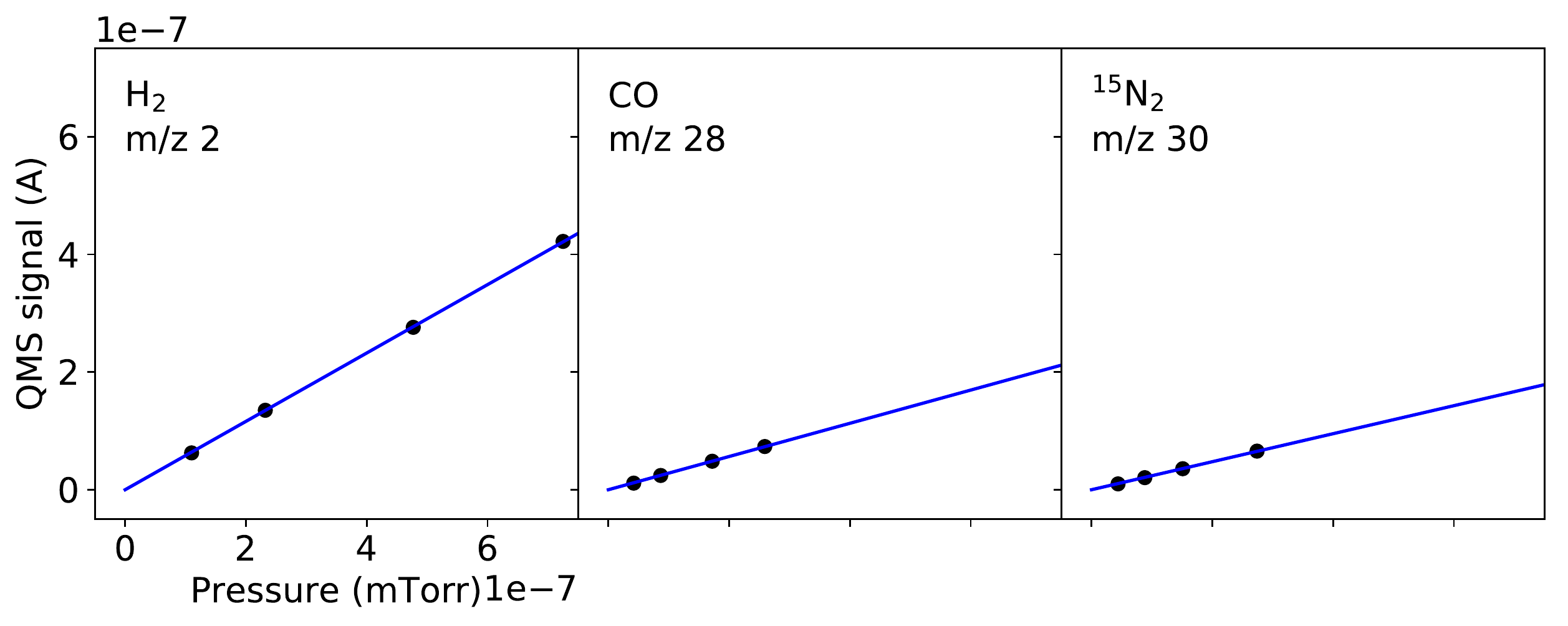}
    \caption{Linear fit (solid blue line) to the measured QMS signal of the main mass fragments of H$_2$ ($m/z$ = 2, left panel), CO ($m/z$ = 28, middle panel), and $^{15}$N$_2$ ($m/z$ = 30, right panel) at different pressures of the pure gases in SPACE TIGER (black dots). Similar fits were determined for the rest of isotopologues used in this work.}
    \label{fig:qms_calib}
\end{figure}

In order to estimate the ice composition from the partial pressures of the different components in the gas mixtures 
we derived a conversion factor for each species that transformed the QMS signal of the ice component main mass fragments into partial pressures. 
To that end, we previously introduced different pressures of the pure gases in the UHV chamber, 
and measured the QMS signal of the species main mass fragment. 
We note that the pressure read by the baratron gauge was corrected by the relative probability of ionization with respect to N$_2$, as indicated by the manufacturer. 
The $^{12}$CO relative probability of ionization was used for the $^{13}$CO and C$^{18}$O isotopologs as an approximation. 
Fig. \ref{fig:qms_calib} shows the linear relation between the corrected gas pressure and the corresponding QMS signal of H$_2$, CO, and $^{15}$N$_2$ molecules.  
In Experiments 4 and 6 the carbon monoxide and  molecular nitrogen molecules shared their main mass fragments ($m/z$ = 28 and $m/z$ = 30, respectively), and alternative mass fragments were used ($m/z$ = 12 for CO and C$^{18}$O, $m/z$ = 14 for N$_2$, and $m/z$ = 15 for $^{15}$N$_2$)

{}


\subsection{Ice thickness}\label{sec:thickness_app}

\begin{deluxetable}{ccccc}
\tablecaption{Summary of the additional experimental simulations.\label{table_exp_app}}
\tablehead{
\colhead{Exp.} & \colhead{Ice comp.}  & \colhead{Thickness} & \colhead{Irradiation} & \colhead{Incident energy}\\
& & \colhead{(ML)$^{a}$} & & \colhead{($\times$ 10$^{18}$ eV)} 
}
\startdata
A1 & CO & 1000 & 2 keV e$^-$ & 1.4 \\
\hline
A2 & CO & 400 & Ly-$\alpha$ & not measured$^c$\\
\hline
C1$-$C7 & CO & 217$-$924$^b$ & blank & \nodata\\
\hline
\enddata
\tablecomments{
$^{a}$ 1 ML = 10$^{15}$ molecules cm$^{-2}$. We assume a 20\% error in the ice thickness in Experiments A1$-$A2 (see Sect. \ref{sec:thickness_app}). 
$^b$ The ice thickness was calculated using Eq. \ref{eqn} from the IR spectra collected in transmission mode.
}
\end{deluxetable}{}

The total thicknesses of the ice samples in Experiments 1$-$14, A1, and A2 were calculated in terms of molecule column densities (molecules cm$^{-2}$). 
Due to the non-linear behavior of the IR absorbance with the species column densities above a certain threshold in the IR spectra collected in reflection-absorption mode, the CO IR feature at $\sim$2140 cm$^{-1}$ could not be used to determine the initial CO ice column density \citep[see, e.g.][]{oberg09}.
As mentioned in Sect. \ref{sec:tpd}, the initial CO ice column density ($N(CO)$) 
was estimated instead from the area under the TPD curve measured with the QMS  ($A_{TPD}(m/z=28)$), while the initial H$_2$ and $^{15}$N$_2$ ice column densities were estimated from $N(CO)$ assuming that the initial ice composition was the same as the composition of the gas mixture used for ice deposition. 
We note that the error made in the approximation of the initial CO ice column density as the CO column density calculated with the TPD curve after processing of the ice samples (below 10\% according to the CO conversion into products in the 2 keV electron irradiation experiments, Sect. \ref{sec:co_conv})  was lower than the uncertainty in the estimation of the CO column density (20\%, see below).  

The area under the CO TPD curve ($A_{TPD}(m/z=28)$) was proportional to the CO ice column density\footnote{Actually, the area under the CO TPD curve is proportional to the absolute number of desorbing CO molecules. If we assume that the ice surface in Experiments 1$-$12 is the same as in the calibration Experiments C1$-$C7, $A_{TPD}(m/z=28)$ is also proportional to 
the CO ice column density.}. 
We calibrated the QMS to extract the proportionality constant $k_{CO}$:

\begin{equation}
    k_{CO} = \frac{A_{TPD}(m/z=28)}{N(CO)}
\end{equation}{}

\begin{figure}
    \centering
    \includegraphics[width=5cm]{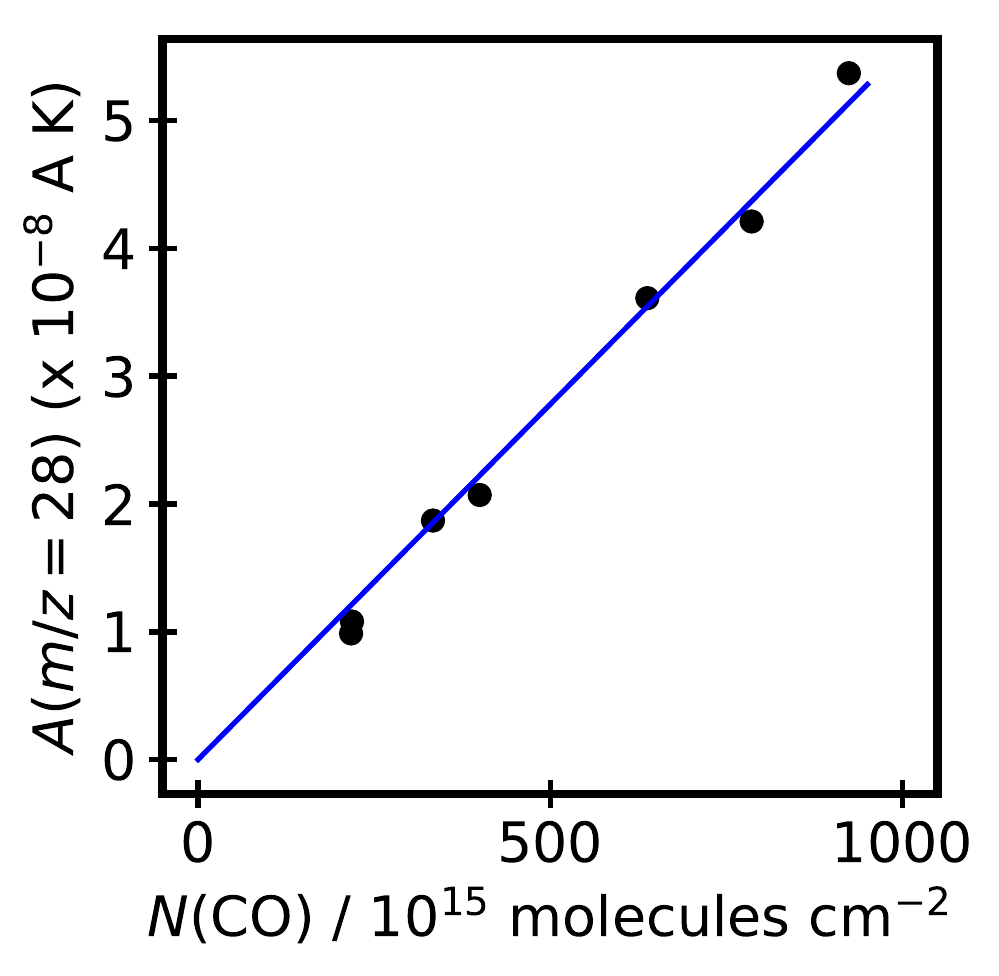}
    \caption{Linear relation (solid blue line) between the area under the TPD curve of a pure CO ice in SPACE TIGER ($A(m/z=28)$) and the CO ice column density $N$(CO), extracted from a series of seven calibration experiments (black dots).}
    \label{fig:CO_calib}
\end{figure}

The same proportionality constant was used for the $^{13}$CO isotopolog in Experiments 5 and 6, as a first approximation.
The constant $k_{CO}$ was calculated from a series of seven calibration experiments in which pure CO ice samples were deposited on top of a CsI IR transparent substrate that allowed collection of the IR spectrum in transmission mode (Experiments C1$-$C7, Table \ref{table_exp_app}). 
The CO ice column density ($N(CO)$) was subsequently estimated from the $\sim$2140 cm$^{-1}$ IR feature using Eq. \ref{eqn}.
%
%
The $\sim$2140 cm$^{-1}$ IR feature absorbance was numerically integrated using the composite Simpson's rule (\textit{integrate.simps} in Python). 
The band strength $A$ of this feature is
1.1 $\times$ 10$^{-17}$ cm mol.$^{-1}$ (Table \ref{table_band}). 
After deposition, a heating rate of 2 K min$^{-1}$ was applied to the pure CO ice until complete sublimation was achieved. 
The area under the CO TPD curve was also calculated with the \textit{integrate.simps} function in Python. 
A linear fit was subsequently performed to the seven experimental data points in Fig. \ref{fig:CO_calib}, using the \textit{curve$\_$fit} function in Python. The proportionality constant $k_{CO}$ was found to be [5.56 $\pm$ 0.11] $\times$ 10$^{-11}$ A K ML$^{-1}$. 
A 20\% uncertainty was assumed due to the 20\% uncertainty of the CO IR band strength \citep{bouilloud15}.

{}

In Experiments 4 and 6 the carbon monoxide and  molecular nitrogen molecules shared their main mass fragments ($m/z$ = 28 and $m/z$ = 30, respectively), and the ice thickness was determined from $A(m/z=12)$, for which a similar calibration was performed. 

On the other hand, the total thickness of the ice samples in Experiments 13 and 14 was calculated from the $^{15}$N$_2$ ice column density ($N(^{15}N_2$)).
The $^{15}$N$_2$ initial column density was calculated from the area $A_{TPD}(m/z=30)$ under the corresponding TPD curve. 
The proportionality constant $k_{^{15}N_2}$ was averaged from the $A_{TPD}(m/z=30)$/${N(^{15}N_2)}$ measured in Experiments 1$-$3, 5, 7, 8, 11, and 12, and had a value of 8.1 $\times$ 10$^{-11}$ A K ML$^{-1}$.

\section{Ice chemistry in the apolar ice layer}

\subsection{Uncertainties in the product conversion yields}\label{sec:quan_app_qms} 


\begin{figure}
    \centering
    \includegraphics[width=9cm]{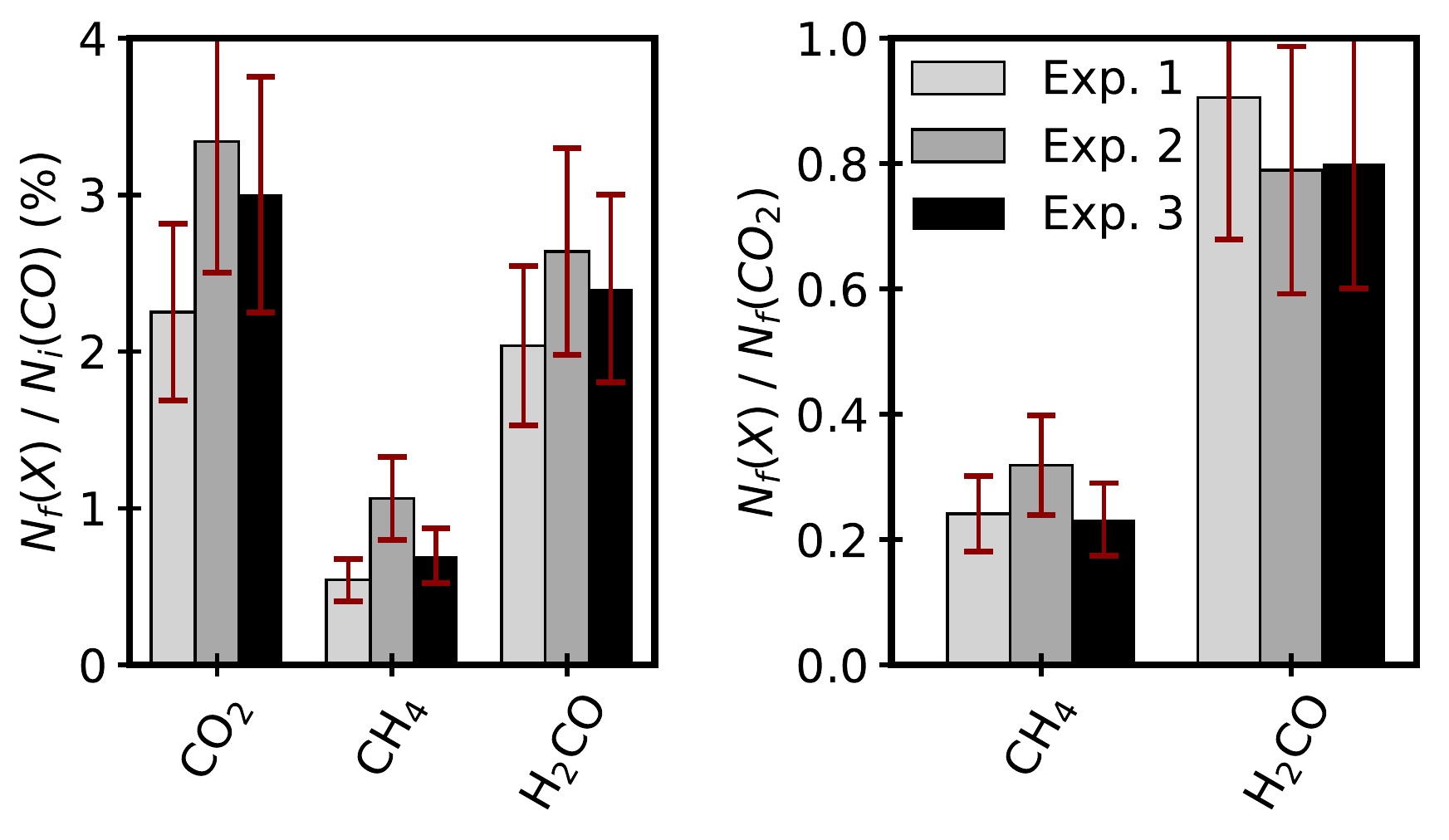}
    \caption{Conversion yields (left panel), and relative formation with respect to CO$_2$ (right panel) of the ice chemistry products with detected IR features in 2 keV electron irradiated H$_2$:CO:$^{15}$N$_2$ ice analogs (Experiments 1$-$3). The  assumed 25\% experimental uncertainty is indicated as red errorbars.} 
    \label{fig:production_e}
\end{figure}{}

The 2 keV electron irradiation of H$_2$:CO:$^{15}$N$_2$ ice analogs in Experiments 1, 2, and 3 were performed under the same conditions, and the same results were thus expected. 
%
The left panel of Fig. \ref{fig:production_e} shows 
the CO$_2$, CH$_4$, and H$_2$CO conversion yields ($N_f(X)$/$N_i(CO)$, Sect. \ref{sec:conv}). 
We found a 25\% experimental uncertainty (indicated as red errorbars) in the product conversion yields measured in Experiments 1$-$3, on top of the systematic uncertainties in the product IR band strengths and the calculated $^{13}$CO/CO fraction, described in Sect. \ref{sec:conv}. 
The systematic uncertainties have the same effect in all experiments, and should not be taken into account when comparing the conversion yields (or product branching ratios) of the same species in different experiments. 
Several sources of errors contributed to this experimental uncertainty. For example, the uncertainty in the integrated $^{13}$CO IR absorbance (due to the low signal-to-noise ratio of this feature), or other day-to-day changes in the precise experimental conditions, such as the noise level in the collected IR spectra.
The right panel of Fig. \ref{fig:production_e} shows the CH$_4$ and H$_2$CO product branching ratios with respect to CO$_2$ in terms of conversion yield ratios. 
The additional $\sim$25\% experimental uncertainty mentioned above was able to account for the differences found between Experiments 1$-$3.
%
%

\subsection{Ice chemistry in analogs of the apolar ice layer}\label{sec:full_spec_app}

\begin{figure*}
    \centering
    \includegraphics[width=13cm]{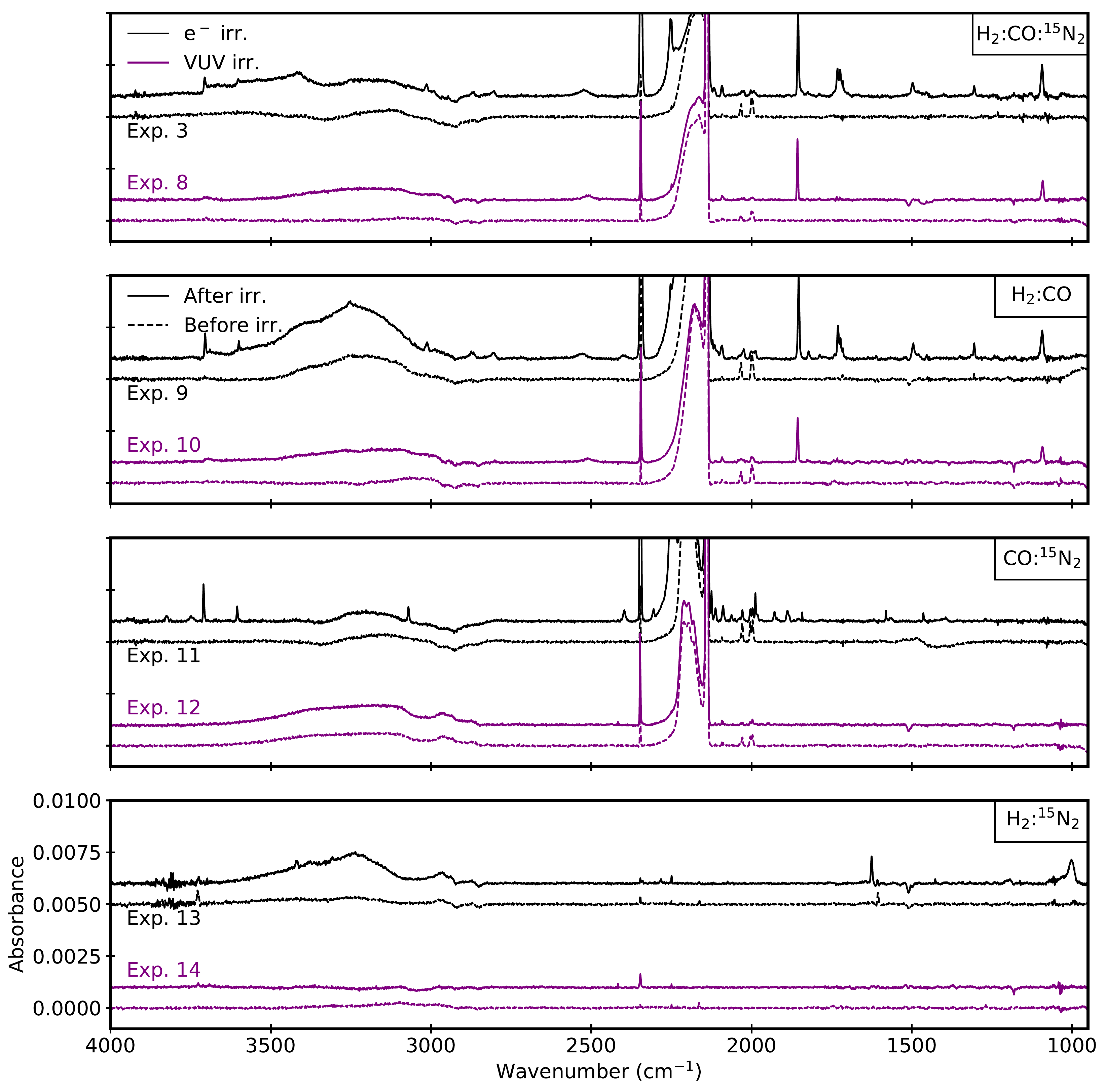}
    \caption{IR spectra obtained before (dashed lines) and after (solid lines) irradiation of a similar incident energy with 2 keV electrons (black) and Ly-$\alpha$ photons (purple) of H$_2$:CO:$^{15}$N$_2$, H$_2$:CO, CO:$^{15}$N$_2$, and H$_2$:$^{15}$N$_2$ ice samples (from top to bottom). 
    The IR spectra are offset for clarity}
    \label{fig:full_spec_app}
\end{figure*}{}

Fig. \ref{fig:full_spec_app} shows the IR spectra collected before and after the energetic processing of the analogs of the apolar ice layer in Experiments 3 and 8$-$12 in the mid-IR range. The difference spectra, along with the identification of the more relevant features are presented in Sect. \ref{sec:h2_co_n2}.
A broad feature above 3000 cm$^{-1}$ was usually observed after irradiation of the different ice samples, regardless of their composition. This feature was probably due to the condensation of H$_2$O molecules on the walls of the IR detector during the experimental simulations. Contribution from background contamination inside the UHV chamber was not expected to be significant. Unfortunately, this prevented us from studying the potential formation of H$_2$O molecules in the processed ice samples. 

\subsection{Ice chemistry of a pure CO ice sample}\label{sec:co_app}

\begin{figure}
    \centering
    \includegraphics[width=12cm]{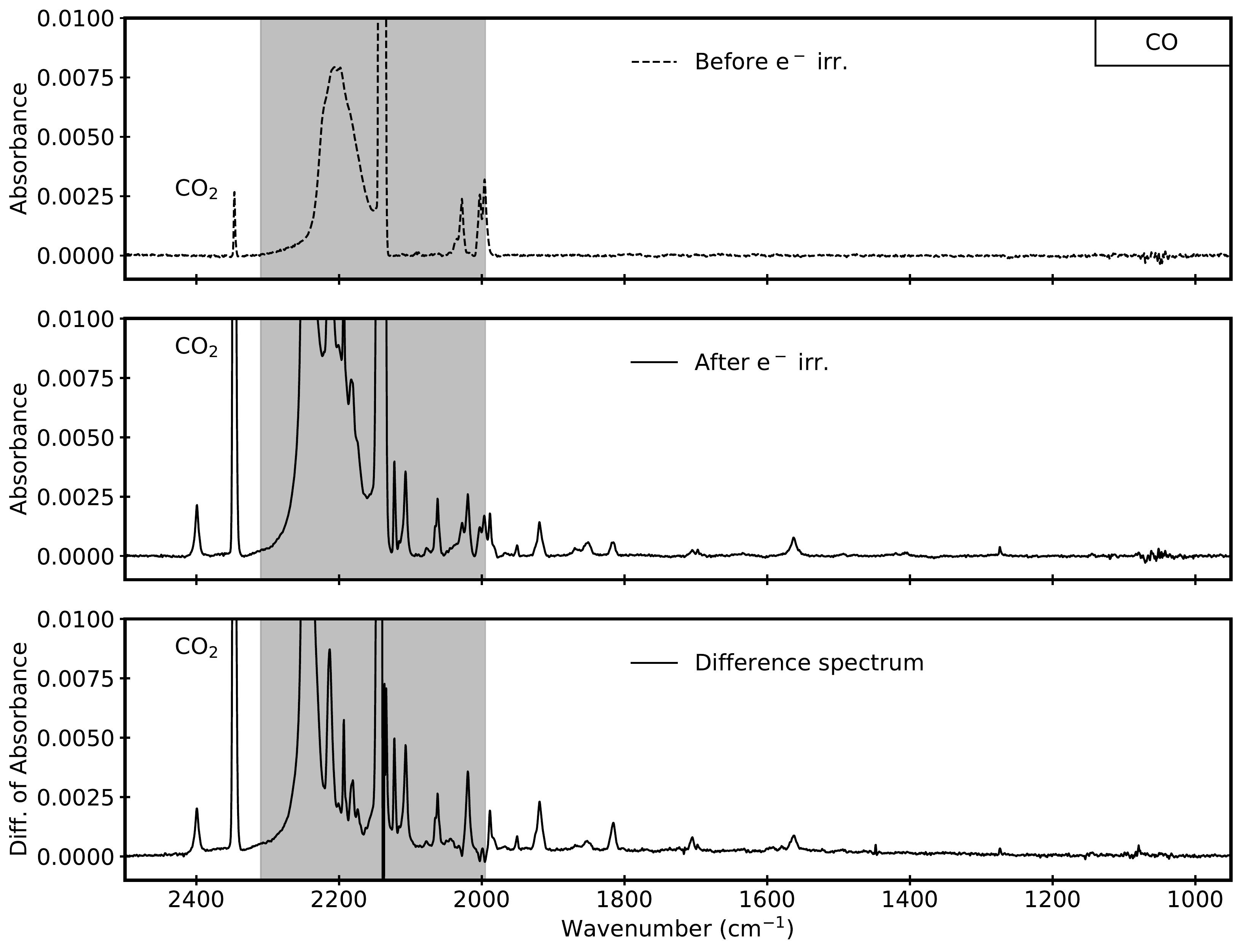}
    \caption{IR spectrum before (top panel), and after (middle panel) a total deposited energy of $\sim$1.3 $\times$ 10$^{18}$ eV  through 2 keV electron irradiation of a 1000 ML pure CO ice (Exp. A1, Table \ref{table_exp_app}), along with the IR difference spectrum (bottom panel). The 2310$-$1995 cm$^{-1}$ region, where the CO and most of the carbon chain oxides IR features are located, is highlighted in every panel.}
    \label{fig:co_ir}
\end{figure}

The top panel of Fig. \ref{fig:co_ir} shows the IR spectrum of a 1000 ML pure CO ice (Exp. A1, Table \ref{table_exp_app}) in the 2500$-$950 cm$^{-1}$ region of the spectrum. 
The CO IR features were confined to the 2310$-$1995 cm$^{-1}$ region, and only a small CO$_2$ feature due to contamination 
was observed outside that region.  
The CO$_2$ contamination represented $\sim$0.01\% of the initial CO molecules, as measured from the IR spectra in transmission mode in Experiments C1$-$C7. 
After a total deposited energy of $\sim$1.3 $\times$ 10$^{18}$ eV through 2 keV electron irradiation, 
the most abundant formed species was CO$_2$ \citep{ioppolo09}.  
A bunch of additional IR features were detected, most of them (but not all) in the 2310$-$1995 cm$^{-1}$ region of the spectrum (middle panel of Fig. \ref{fig:co_ir}), corresponding to carbon chain oxides including C$_3$O, C$_3$O$_2$, C$_4$O, C$_4$O$_2$, C$_5$O, C$_5$O$_2$, C$_7$O, and C$_7$O$_2$.
A C$_2$O IR feature was detected at the edge of this region \citep[at 1989 cm$^{-1}$,][]{palumbo08}, blended with a CO IR band.
A complete analysis of the formation of carbon chain oxides upon energetic processing of CO-bearing ice samples is presented in \citet{sicilia12}, and it is beyond the scope of this paper. 

In addition, the difference spectrum (bottom panel of Fig. \ref{fig:co_ir}) shows that the 2140 cm$^{-1}$ CO feature does not decrease, but increases its intensity upon energetic processing of the pure CO ice. A similar increase is observed after energetic processing of H$_2$:CO:$^{15}$N$_2$, H$_2$:CO, and CO:$^{15}$N$_2$ ice samples (Fig. \ref{fig:ir_bi}). 
This is probably due to a change in the C=O stretching IR band strength as a result of structural changes taking place in the ice sample, combined with the limited chemistry undergone by the CO molecules.

{}

\subsection{Product identification in different isotopically labeled ice mixtures}\label{sec:products_app}

\begin{figure*}
\gridline{\fig{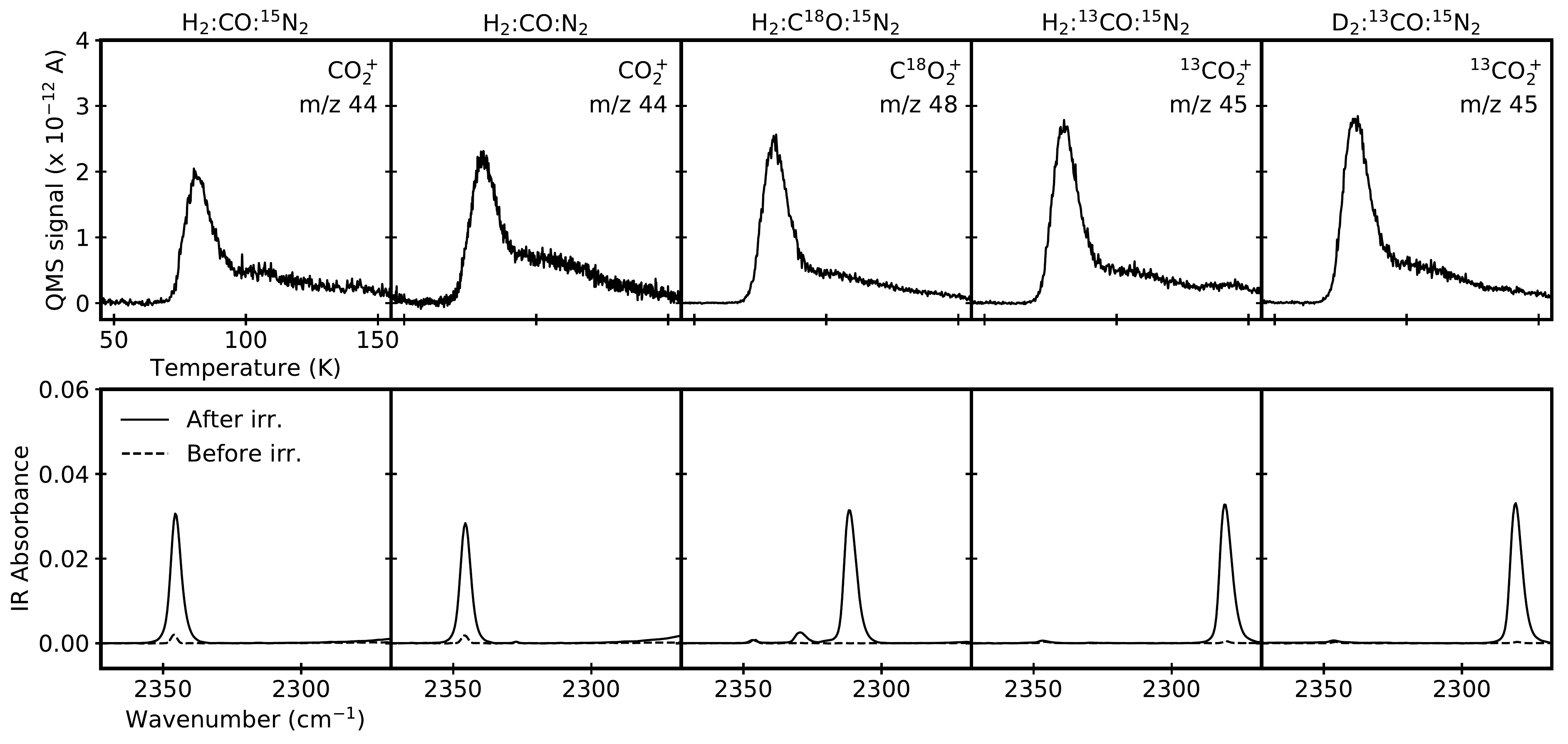}{0.8\textwidth}{}}
\gridline{\fig{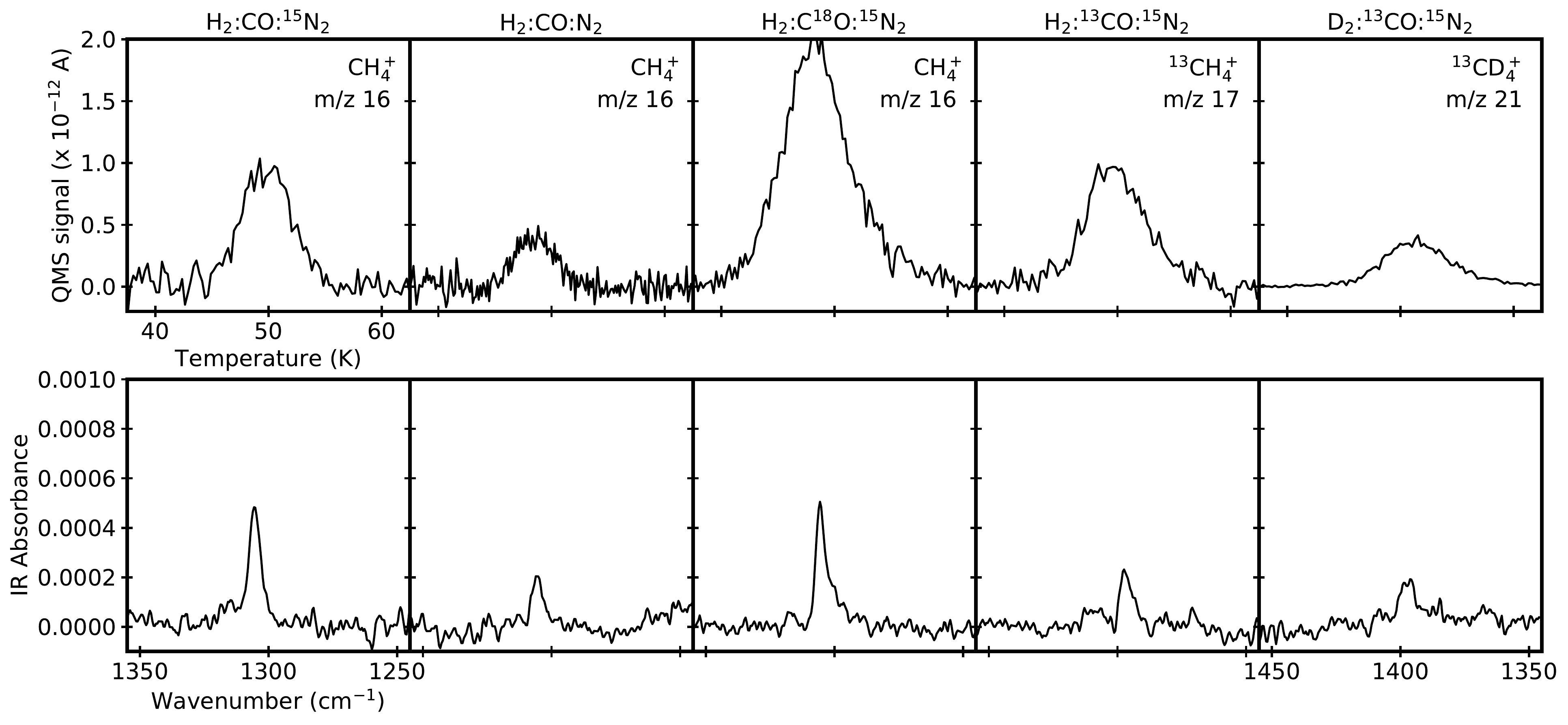}{0.8\textwidth}{}}
\gridline{\fig{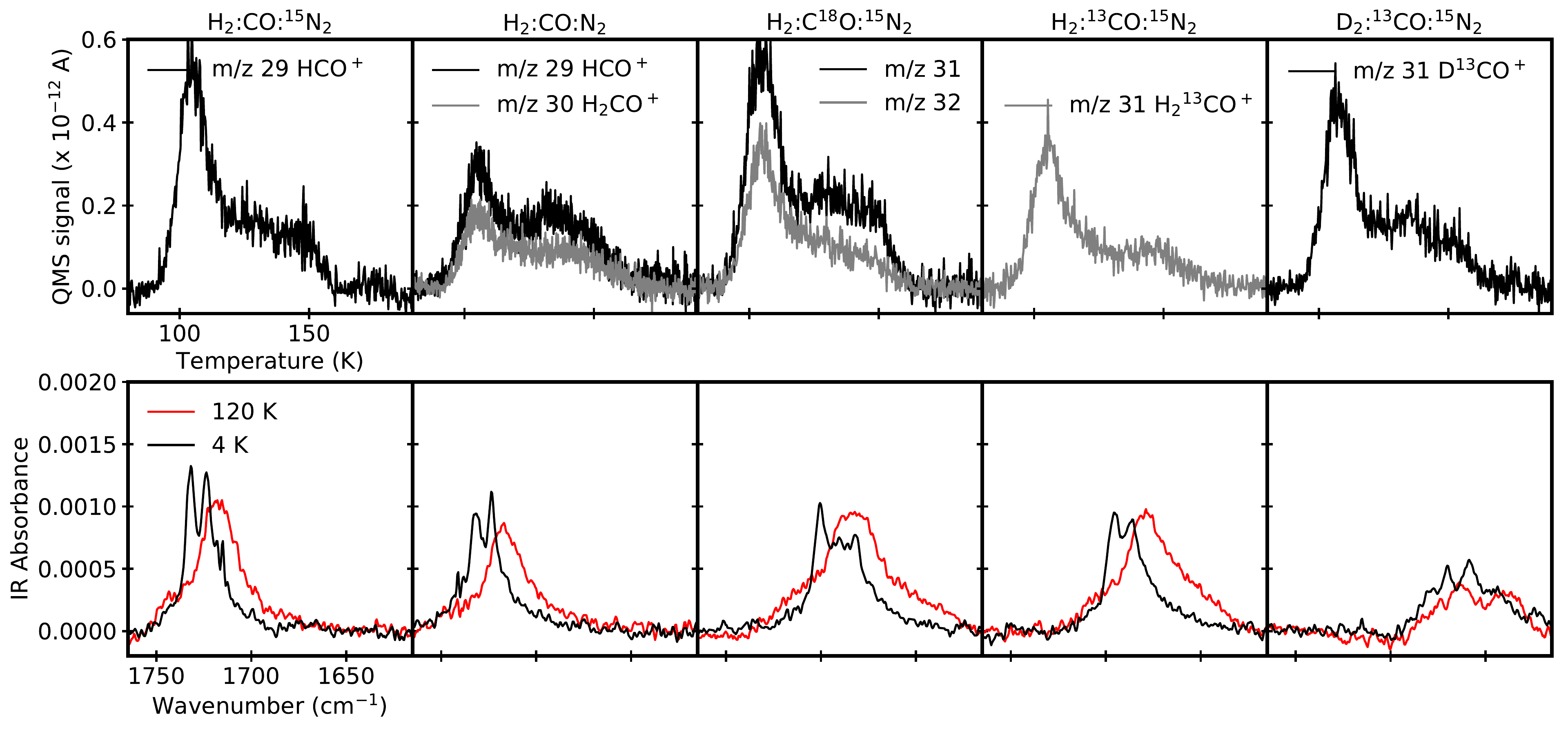}{0.8\textwidth}{}}
\caption{TPD curves and IR features of the CO$_2$, CH$_4$, and H$_2$CO (from top to bottom) produced upon 2 keV electron irradiation of different isotopically labeled three-component ice samples, from left to right: H$_2$:CO:$^{15}$N$_2$, H$_2$:CO:N$_2$, H$_2$:C$^{18}$O:$^{15}$N$_2$, H$_2$:$^{13}$CO:$^{15}$N$_2$, and D$_2$:$^{13}$CO:$^{15}$N$_2$.
The H$_2$CO TPD curves are shown for both the main mass fragment (black) and the molecular mass fragment (gray). The H$_2$CO IR feature observed at 4.3 K (black) is shown along with that detected at 120 K (red). 
\label{fig:ir_det}}
\end{figure*}

In order to confirm the assignments of the thermal desorption peaks observed in the left panels of Fig. \ref{fig:tpd_bi}, we irradiated four additional ice analogs composed by different combinations of isotopically labeled H$_2$, CO, and N$_2$ molecules with 2 keV electrons (Experiments 4$-$7 in Table \ref{table_exp}). 
The positions of the IR bands detected after irradiation of the ice samples, and the mass fragments showing thermal desorption during the TPD of the processed ices were shifted according to the particular isotopic composition of the products in the different ice samples. This is shown in Fig. \ref{fig:hnco_det} for HNCO, H$_2$C$_2$O and C$_2$O (only TPD curves), and in Fig.  \ref{fig:ir_det} for CO$_2$, H$_2$CO, and CH$_4$. 
For example, the C=O stretching IR band corresponding to CO$_2$ shifted from 2345 cm$^{-1}$ to 2310 cm$^{-1}$ for C$^{18}$O$_2$, and 2280 cm$^{-1}$ for $^{13}$CO$_2$; while the thermal desorption was observed for the mass fragments $m/z$ = 44, 48, and 45, respectively. 

The double-peaked H$_2$CO C=O stretching IR band at 1725 cm$^{-1}$ was also redshifted for H$_2$C$^{18}$O, H$_2$$^{13}$CO, and D$_2$$^{13}$CO,  
However, the double peak structure was not observed at higher temperatures (only one peak was detected, see Fig. \ref{fig:ir_det}). 
This double peak structure could be thus due to the apolar ice matrix in which the produced H$_2$CO molecules were embedded, since the majority of the H$_2$, CO, and N$_2$ molecules conforming this apolar ice matrix had already desorbed at higher temperatures. 
In addition, as mentioned in Sect. \ref{sec:h2_co_n2}, the thermal desorption of H$_2$CO presented two desorption peaks. 
We speculate that the second desorption peak could be due to the formation of H$_2$CO dimers with a higher binding energy, since the H$_2$CO IR feature was detected at temperatures higher than the multilayer H$_2$CO thermal desorption peak temperature. However, the two desorption peaks were only observed for the mass fragments corresponding to HCO$^+$ and H$_2$CO$^+$ (and the corresponding isotopologs), but not to the (H$_2$CO)$_2$ dimers (not shown in Fig. \ref{fig:ir_det}).

\end{document}